# Giant persistent photoconductivity in monolayer $MoS_2$ field-effect transistors

A. George[1, 2], M. V. Fistul[3,4,5], M. Gruenewald[6], D. Kaiser[1], T. Lehnert[7], R. Mupparapu[2,8], C. Neumann[1], U. Hübner[9], M. Schaal[6], N. Masurkar[10], A. L. M. Reddy[10], I. Staude[2,8], U. Kaiser[7], T. Fritz[6], A. Turchanin[1, 2*]

*[1]Friedrich Schiller University Jena, Institute of Physical Chemistry, 07743 Jena, Germany*
*[2]Abbe Centre of Photonics, 07743 Jena, Germany*
*[3]Institute for Basic Science (IBS), Center for Theoretical Physics of Complex Systems, 34126 Daejeon, Republic of Korea*
*[4]Ruhr-University Bochum, Theoretische Physik III, 44801 Bochum, Germany*
*[5]National University of Science and Technology (MISiS), 119049 Moscow, Russia*
*[6]Friedrich Schiller University Jena, Institute of Solid State Physics, 07743 Jena, Germany*
*[7]Ulm University, Central Facility of Electron Microscopy, Electron Microscopy Group of Materials Science, 89081 Ulm, Germany*
*[8]Friedrich Schiller University Jena, Institute of Applied Physics, 07745 Jena, Germany*
*[9]Leibniz Institute of Photonic Technology, 07745 Jena, Germany*
*[10]Wayne State University, Department of Mechanical Engineering, 48202 Detroit, USA*

e-mail: andrey.turchanin@uni-jena.de
Tel.:+49-3641-948370
Fax: +49-3641-948302

**ABSTRACT**

Monolayer transition metal dichalcogenides (TMD) have numerous potential applications in ultrathin electronics and photonics. The exposure of TMD based devices to light generates photo-carriers resulting in an enhanced conductivity, which can be effectively used, e.g., in photodetectors. If the photo-enhanced conductivity persists after removal of the irradiation, the effect is known as persistent photoconductivity (PPC). Here we show that ultraviolet light ($\lambda$ = 365 nm) exposure induces an extremely long living *giant PPC* (GPPC) in monolayer $MoS_2$ (ML-$MoS_2$) field-effect transistors (FET) with a time constant of ~30 days. Furthermore, this effect leads to a large enhancement of the conductivity up to a factor of $10^7$. In contrast to previous studies in which the origin of the PPC was attributed to extrinsic reasons such as trapped charges in the substrate or adsorbates, we unambiguously show that the GPPC arises mainly from the intrinsic properties of ML-$MoS_2$ such as lattice defects that induce a large amount of localized states in the forbidden gap. This finding is supported by a detailed experimental and theoretical study of the electric transport in TMD based FETs as well as by characterization of ML-$MoS_2$ with scanning tunneling spectroscopy, high-resolution transmission electron microscopy and photoluminescence measurements. The obtained results provide a basis towards the defect-based engineering of the electronic and optical properties of TMDs for device applications.

## TOC IMAGE

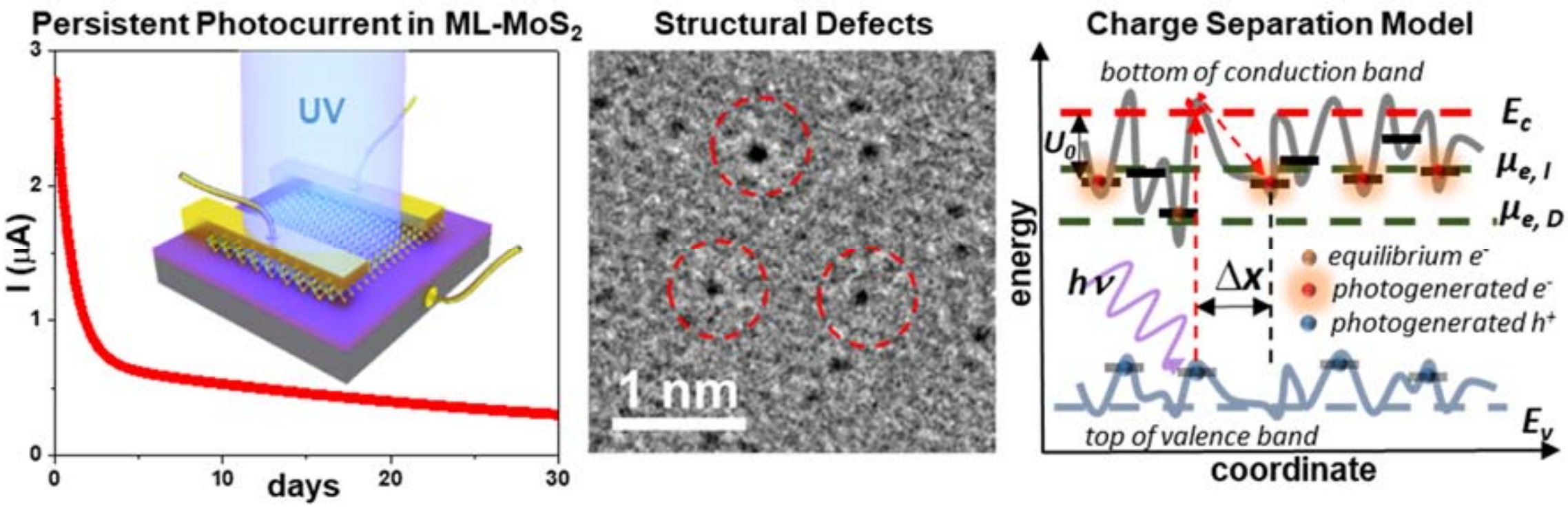

## TOC TEXT

We report on a giant persistent photoconductivity in monolayers of $MoS_2$ with a time constant of more than 30 days after irradiation with UV light. Our detailed microscopy study demonstrates that the main reason of this phenomenon are the intrinsic structural defects resulting in strong spatial variation of the monolayer band structure.

Persistent photoconductivity (PPC) has long been studied in amorphous as well as highly-compensated wide-bandgap bulk semiconductors and was attributed to the presence of large spatial fluctuations of the potential energy of charge carriers (electrons and holes).[1, 2, 3] In case of transition metal dichalcogenides (TMD), the PPC effect with the respective time constant, $\tau$, of $10^2$ -$10^4$ s was reported for monolayer $MoS_2$ (ML-$MoS_2$) after their irradiation with visible light at room temperature (RT).[4, 5, 6, 7] An even higher $\tau \sim 10^6$ s was observed for few layers of $MoS_2$ after UV irradiation ($\lambda$ = 254 nm).[8] In these studies, the PPC effect was related to the charge traps caused by inhomogeneities either in the substrate[4, 6, 9] or in the adsorbates[8, 9] on the TMD surface. Here we show that the long living photo-generated charge carriers may also originate from intrinsic, material specific lattice defects resulting in a prolonged recombination time of photo-generated carriers.

We demonstrate the extremely long living *giant persistent photoconductivity* (GPPC) in field effect transistors (FETs) fabricated from single crystalline ML-$MoS_2$ grown by chemical vapor deposition[10, 11] (CVD) after their exposure to UV light ($\lambda$ = 365 nm), Fig. 1. At RT the photo-generated charge carriers lead to an increase of the conductivity by a factor of up to ~$10^7$, which depends on both the applied gate voltage ($V_g$) and the irradiation intensity (Fig. 1b). The high conductivity state persists for a long time with a time constant of ~30 days ($3\times10^6$ s) at $V_g$ = 0 V (Fig. 1c). We explain these experimental findings with a model considering the presence of *large spatial fluctuations of the potential energy* of carriers (electrons and holes) in the ML-$MoS_2$. These fluctuations lead to a spatial separation of photo-generated carriers, as electrons (holes) concentrate in the minima (maxima) of random potential energy landscape (see schematic in Fig. 1d) resulting in a giant increase of their recombination time.[12, 13, 14] Transport of these photo-generated carriers displays two regimes:

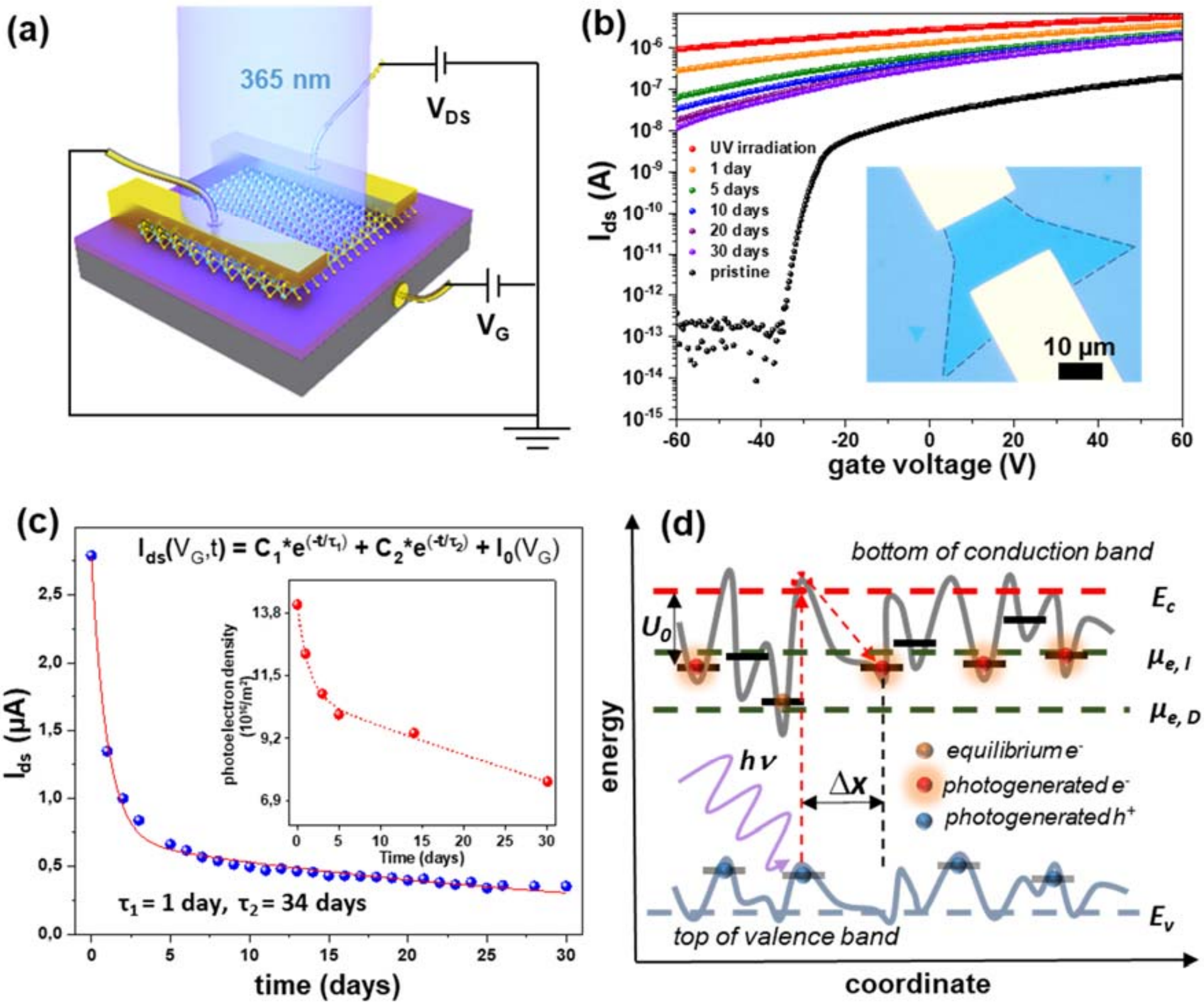

**Figure 1.** (a) Schematic diagram of the $MoS_2$-FET device and experimental setup. (b) Experimental observation of GPPC in a $MoS_2$-FET device. The black curve represents the transfer characteristics of the $MoS_2$ device before UV irradiation, the red curve represents the transfer characteristics immediately after UV irradiation ($\lambda$ = 365 nm) for 5 min with an intensity of ~30 mW/cm$^2$. The colored curves represent the time-dependent transfer curves after UV irradiation. The measurements continued up to 30 days (complete data set is provided in Fig. S4). The inset shows an optical microscopy image of the $MoS_2$-FET device. (c) The decay of the drain current with time at $V_g$ = 0 V. The experimental data were fitted (red curve) using a two-stage exponential decay function to extract the GPPC time constants. Calculated decay of the photoelectron concentration with time is shown in the inset. (d) Schematic representation of large spatial fluctuations of the potential energy of carriers in a $MoS_2$ monolayer. The incident photons excite electrons from the valence band to the conduction band resulting in their spatial separation $\Delta x$. $\mu_{e,D}$ and $\mu_{e,I}$ are quasi-Fermi energy levels corresponding to equilibrium electrons and photo-generated electrons, respectively.

(i) the thermal activation regime at room temperature (RT) and the variable-range hopping regime at low temperatures (LT). Carrying out a quantitative analysis of the transfer characteristics in both regimes, we extract such parameters of the random potential energy landscape as the characteristic amplitude and the correlation radius as well as the variation of the concentration of photo-generated carriers with time (inset in Fig. 1c) and UV irradiation intensity (Fig. 2e, f). Large fluctuations of the potential energy result in a substantial concentration of strongly localized states in the forbidden energy gap. By performing scanning tunneling spectroscopy (STS) we experimentally confirm the presence of such localized states in ML-$MoS_2$. Atomically-resolved transmission electron microscopy (TEM) enables us to correlate these findings with the density of the point defects in the samples.

### Transport measurements of $MoS_2$-FETs and observation of GPPC

In Fig. 1a a schematic representation of the experimental setup for the transport measurements is shown. As grown ML-$MoS_2$ were characterized by optical microscopy, Raman spectroscopy, and atomic force microscopy (Figs. S1-2). Afterwards, the FETs were fabricated using e-beam lithography. An optical microscopy image of a typical device is shown in the inset of Fig. 1b. We measured the transfer characteristics of these devices, i.e. the drain-source current, $I_{ds}$, versus $V_g$ before and after UV irradiation ($\lambda$ = 365 nm, intensity of ~30 mW/cm$^2$ for 5 minutes). All measurements were performed in a high vacuum (~$10^{-6}$ mbar) and under dark conditions; these data are shown in Fig. 1b. Note that we studied a possible effect of the UV irradiation induced damage of the ML-$MoS_2$ by conducting Raman spectroscopy before and after the irradiation and did not find any noticeable changes in the spectra (see Fig. S3).

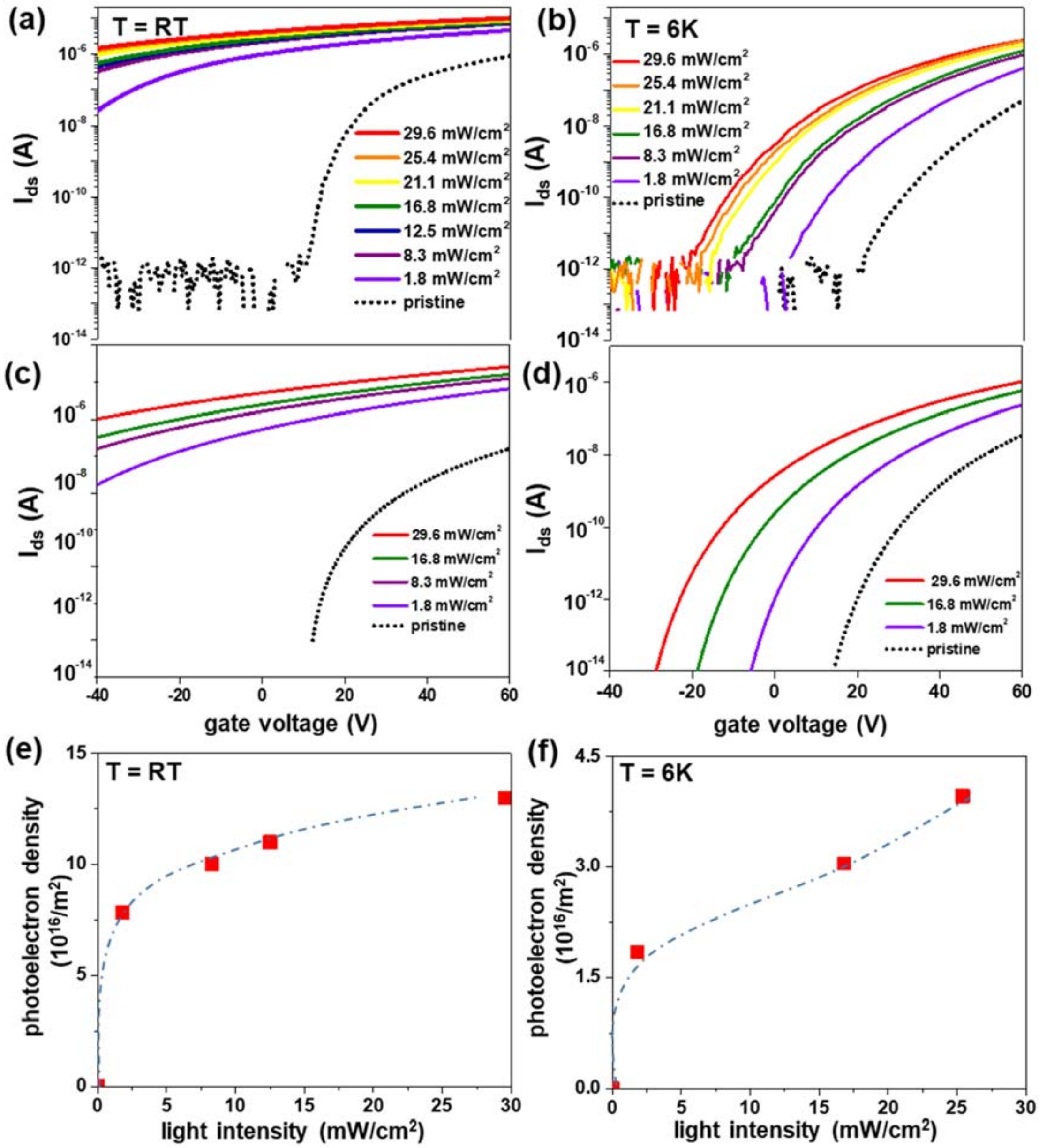

**Figure 2**. Transfer characteristics of a $MoS_2$ FET at the pristine condition and after UV irradiation with various intensities. The irradiation time for each intensity was 5 min, and all transfer curves were recorded in dark immediately after irradiation. Experimental transfer curves at RT (a) and at 6 K (b). Theoretically calculated transfer curves RT (c) and at 6 K (d). Calculated photoelectron density as a function of the irradiation intensity at RT (e) and at 6 K (f). The dashed lines in (e) and (f) are presented as a guide for the eye. The GPPC at a function of time for this device is presented in Fig. S4.

From the transfer characteristics of the pristine device (black line in the Fig. 1b) we estimate a field effect mobility of 1.5 $cm^2/Vs$, which is a typical value for CVD grown ML-$MoS_2$.[11, 15] Directly after UV irradiation, we observe a very strong enhancement in the $I_{ds}$ (see red line in Fig. 1b) of up to $\frac{I_{irr}}{I_{non-irr}} \approx 10^7$ at $V_g$ = - 40 V, which is close to the device threshold voltage. We found that the GPPC persists even for days at RT, albeit decaying in its strength over time (transfer characteristics recorded between 1 and 30 days are shown in Fig. 1b; Fig. S4 shows the full data set). During these measurements the devices were always kept in high vacuum and under dark conditions. The obtained decay of $I_{ds}$ over time is shown in Fig. 1c at $V_g$ = 0 V. These data can be described using a two-stage exponential decay function. In the initial stage, the $I_{ds}$ decays with a time constant of $\tau_1 \approx 1$ day, whereas in the following stage, the GPPC relaxation slows down yielding a time constant of $\tau_2 \approx 34$ days. Similar values were obtained for more than 10 devices made of ML-$MoS_2$ synthesized in different CVD experiments (see, e.g., Fig. S5). Note that after a few months, the transfer characteristics of the devices completely recover to their pristine state before irradiation (Fig. S6a). We also found that vacuum annealing (~$10^{-2}$ mbar) at 170 °C results in a significantly faster decrease of the persistent photocurrent at RT (see SI p. S6 and Fig. S6b), which agrees well with the thermally enhanced recombination of the photo-generated carriers. Similar as after their recombination at RT, a subsequent UV irradiation of the device induces the initially observed GPPC effect.

To investigate the GPPC in detail, we measured the transfer characteristics of the $MoS_2$-FETs at RT and low temperature (LT, 6 K) after UV irradiation of varying intensities between 2 and 30 $mW/cm^2$ for 5 min as shown in Fig. 2a and 2b, respectively. In both cases, we observed an enhancement of the $I_{ds}$ with increasing UV intensity. At LT, the absolute values of $I_{ds}$ were lower than at RT for all $V_g$, which we attribute to the variable range hopping type of transport caused by the strong localization of the charge carriers.[16, 17, 18]

**Model of the photo-induced charge transport in the presence of spatial fluctuations of the band structure**

To rationalize the experimental observations of the GPPC in $MoS_2$-FETs, we apply the following model. We assume that the crystal lattice defects and lattice strain lead to large spatial fluctuations of the potential energy of carriers in the ML-$MoS_2$, predominantly because of the deep lying traps in the proximity of the conduction band (CB) / valence band (VB) edges as schematically depicted in Fig. 1d. We describe these fluctuations effectively by a coordinate dependent potential with zero mean value, $U(\boldsymbol{r}) = U_0 f(\boldsymbol{r}/r_{corr})$, where $U_0$ is the typical amplitude of the random potential and in our case $U_0 \gg k_B T$, $T$ is the temperature, $k_B$ is the Boltzmann constant, and $r_{corr}$ is the correlation radius.[12]

By using such a model, we carry out a quantitative analysis of the transfer characteristics in two regimes. At RT, the conductivity of the ML-$MoS_2$ is defined by thermally induced activation of localized electrons surpassing the percolation level (the mobility edge), $E_P$. The $E_P$ is the characteristic energy above which electrons are delocalized, i.e. propagate along a conduction channel experiencing only weak scattering[12, 13]. In this regime, the concentration of delocalized electrons depends strongly on the temperature as well as $V_g$ and thus allows one to tune the conductivity (see dotted line in Fig. 2a). The conductivity in the *thermal activation regime* is obtained as

$$\sigma_{ph}^{RT} = \sigma_0 \exp(-\frac{E_P - \mu_e}{k_B T}), \tag{1}$$

where $\mu_e$ is the quasi-Fermi energy level considering both equilibrium electrons ($\mu_{e,D}$), i.e. the electrons which are present in the absence of irradiation, and non-equilibrium, photo-generated electrons ($\mu_{e,l}$) (see Fig. 1d), and $\sigma_0$ is a coefficient which depends only weakly on both $V_g$ and the intensity of the UV radiation. In strongly disordered materials with many localized states, $\mu_e$ is determined by the concentration of electrons according to[12, 19]

$$n^{eq}(V_g) + n^{non-eq}(J) = n_0 \int_{-\infty}^{\mu_e} g(E)dE, \quad (2)$$

where $n^{eq}$ is the concentration of equilibrium electrons given by $n^{eq}(V_g) = {}^{\varepsilon_r \varepsilon_0 V_g}/_{ed}$, where $\varepsilon_r$ = 3.9 and $d$ = 300 nm are the relative dielectric constant and the thickness of the $SiO_2$ gate insulator, respectively, $\varepsilon_0$ is the vacuum permittivity, $e$ is the electron charge, $n^{non-eq}$ is the concentration of non-equilibrium electrons, $J$ is the intensity of UV irradiation, $n_0$ is the maximum possible concentration of electrons in the CB and $g(E)$ is the density of states. The density of localized states in the bandgap is approximated by [13, 20]

$$g(E) = \frac{2}{U_0}\exp\left(-\frac{2\left|E - E_{CB}^0\right|}{U_0}\right), \; E \le E_{CB}^0, \; \left|E - E_{CB}^0\right| >> U_0, \quad (3)$$

where $E$ is the energy of localized states, and $E_{CB}^0$ is the bottom of the CB in the absence of spatial fluctuations. Note that for disordered two-dimensional (2D) materials the percolation level is given as $E_p = E_{CB}^0$.[13] The calculated transfer characteristic for the pristine, i.e. non-irradiated sample (see dotted line in Fig. 2b), are in a good agreement with the experimental data (dotted line in Fig. 2a) giving the values of $U_0 \cong 0.18\, eV$ and $n_0 \cong 2.7 \times 10^{18}\, m^{-2}$. Since the recombination time of photo excited carriers is expressed as $\tau = \tau_0 \exp\left[\frac{E_a}{k_B T}\right]$, where the activation energy $E_a \approx 2U_0$, in the limit of $U_0 \gg k_B T$ one can expect the GPPC effect.

At LT, a transition to Mott's *variable-range hopping regime* is observed, and for 2D disordered semiconductor materials the conductivity is expressed by[13]

$$\sigma_{ph}^{LT} = \sigma_0 \exp\left[-\left(\frac{T_0}{T}\right)^{1/3}\right], \quad (4)$$

with $T_0 = \frac{13.8}{k_B\, g(\mu_e) r_{corr}^2}$. Using (4) to fit the experimental transfer characteristics for the pristine sample (the dotted lines in Fig. 2b,d), we extract the correlation radius of the random potential $U(\boldsymbol{r})$, $r_{corr}$ = 5 nm. Now, by fitting our experimental transfer data at different UV irradiation doses, $I_{ds}^{RT}(V_g) = V_{ds}\sigma_{ph}^{RT}(V_g)$ and $I_{ds}^{LT}(V_g) = V_{ds}\sigma_{ph}^{LT}(V_g)$, to the theoretical model (see Fig. 2c and 2d) we calculate the variation of the concentration photo-generated electrons, $n^{non\text{-}eq}(J, t)$, with time at RT (inset Fig. 1c) and with UV intensity at RT and LT (see Fig. 2e, f). In agreement with our experiment, the decay of the long living photo-generated electrons with time shows a bi-exponential decay process (see Fig. 1c). We attribute the two different exponents to the variation of the spatial separation between photo-generated electrons and holes, i.e. a part of photo-generated electrons and holes are localized in close proximity and thus recombine faster, whereas another part is separated at longer distances further away from each other and therefore recombine slower.[14] Furthermore, the previously reported photogating[4, 6, 7] effect may also contribute to the faster exponent. Summarizing this part, we conclude that our experimental transport data and the theoretical analysis suggest large spatial fluctuations of the band structures in $MoS_2$-FETs, with the minima of the CB and maxima of the VB serving as trap sites for the photo-generated carriers.

As we show next, our spectroscopy and microscopy study enables us to identify atomic vacancies[21, 22, 23] and strain[24] in the $MoS_2$ monolayers as main reasons for the spatial variation of the band structure responsible for the observed GPPC effect. On the other hand, such extrinsic sources as adsorbates on the FET channel[6] or trapped charged at the ML-$MoS_2$/substrate interface and photogating[4, 6, 7] play a secondary role here (see SI Part 2 and Fig. S7-10 for details of this study).

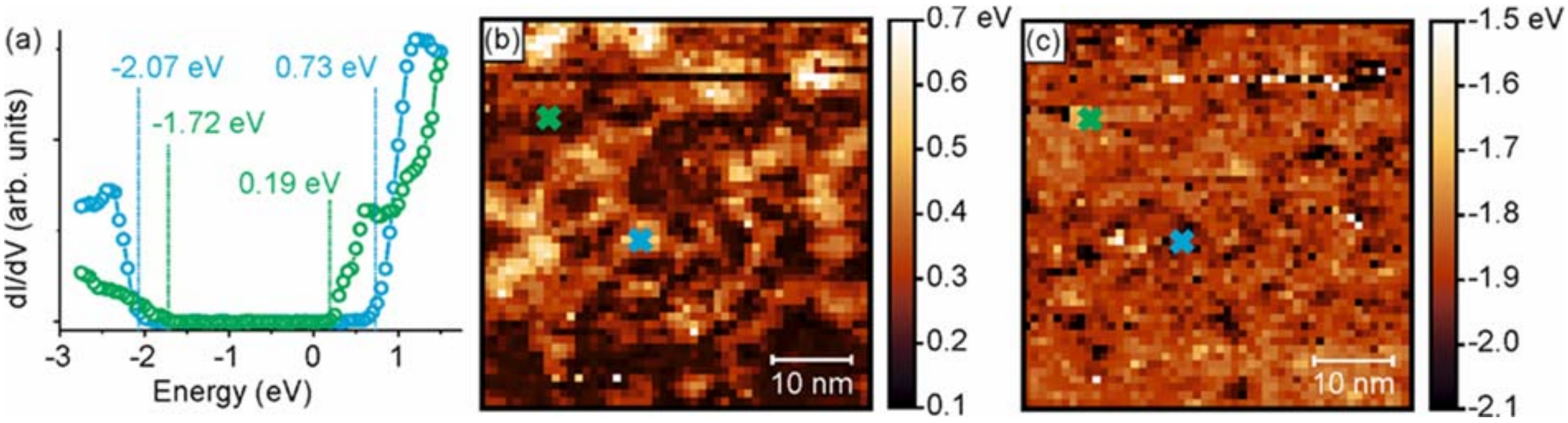

**Figure 3**. (a) STS data obtained at two different positions on $MoS_2$/hBN/Pt(111) showing the onset values determined from the trapped states below the CB and above the VB. STS maps (50 nm × 50 nm) visualizing the spatial distribution of the band structure inhomogeneity are shown in (b) and (c), respectively. The positions where the blue and green curves shown in (a) were obtained, are marked with crosses. The corresponding topography map is provided in Fig. S8.

**Visualization of spatial fluctuations in the band structure by STS**

To visualize the presence of spatial inhomogeneities in the band structure of ML-$MoS_2$, we performed scanning tunneling spectroscopy (STS) at LT (1.1 K). To this end, CVD grown ML-$MoS_2$ was transferred onto a Pt(111) single crystal passivated by a monolayer of hexagonal boron nitride (h-BN) (see SI part 1, Fig. S11 for details), and the derivative of the tunneling current, dI/dV, was measured, which is considered to be proportional to the local density of states (LDOS).[25] Indeed, by STS we observe a spatial dependency of the dI/dV curves from which two extremes are plotted in Fig. 3a. While both curves essentially show the bandgap in ML-MoS2, the green curve additionally reveals features within the band gap, which we identify as deep lying trap states. By evaluating the respective onset of the VB and CB regions we find that bandgap variations are mainly caused by the trap states nearby the CB. As can be seen from Fig. 3b, these trap states spatially form patches with dimensions of about 5-10 nm and show the energy variation of $\sim 0.25\ eV$. Both values are in a very good agreement with the respective parameters $U_0$ and $r_{corr}$ obtained from the theoretical analysis of the transport data. The variation of the band structure in the VB region is less pronounced, as can be seen from Fig. 3c. Note that the topography of the ML-$MoS_2$ obtained by scanning tunneling microscopy (STM) reveals some corrugations (Fig. S12). However, these corrugations do not correlate with the spatial distribution of the trap states[26] (compare Fig. 3b,c and Fig. S13). Therewith we conclude that the apparent monolayer roughness obtained by STM is caused by its interaction with the substrate playing a minor role in the observed fluctuations on the CB and VB. We expect the intrinsic structure to be responsible for that.

**Correlation between spatial inhomogeneity of the band structure with lattice defects and strain**

In order to study the origin of the observed band structure fluctuations in the ML-$MoS_2$, we performed the structural study using the aberration corrected high-resolution transmission electron microscopy (HRTEM).[27] A representative unprocessed image of the ML-$MoS_2$ is shown in Fig. 4a, which clearly demonstrates a presence of the sulfur vacancies.[28, 29] In Fig. 4b, the same image is shown after Fourier filtering of the ML-MoS2 lattice frequency. This procedure facilities direct counting of the vacancies, as they can be recognized as black dots. In this way, we obtain a total concentration of the vacancies in ML-MoS2 of 0.79(6) vac/$nm^2$. Note that for this evaluation only clean areas of the sample were analyzed and an effect of the electron beam induced damage on the ML-MoS2 upon imaging was eliminated (see SI Part 3 and Fig. S14 for details). As the differentiation between single (S1) and double (S2) sulfur vacancies is not possible by counting the black dots in Fig. 4b (S2 vacancies are emphasized in the inserts of Figs. 4a-b with red cycles), the contrast of each vacancy was additionally analysed in the unprocessed HRTEM images (see SI Part 3 and Fig. S15 for details). As a result, the concentration of the $S_2$ vacancies was found to be 0.067(2) vac/$nm^2$, which corresponds to ~8.5 % of the total concentration. Summarizing these results, we ascribe the point detects in ML-$MoS_2$ ($S_1$ and $S_2$ vacancies), their concentration and spatial distribution to the observed fluctuation of the band structure presented in the previous section.

As ML-$MoS_2$ grown by the CVD method is known to build up biaxial strain during the cooling step due to mismatch of the thermal expansion coefficients with the underlying $SiO_2$ substrate,[24] this lattice strain can also cause the band structure fluctuations,[30] which contribute to the observed photoconductivity in the $MoS_2$-FETs. From the frequencies of Raman peaks of in-plane and out-of-plane phonon modes experimentally measured on as-grown $MoS_2$ flakes, we estimate that our CVD grown $MoS_2$ flakes contain a biaxial strain of 0.34±0.08 % (see SI Part 4) relative to the $MoS_2$ flakes exfoliated from the bulk crystal assumed to have zero strain.[24] Most probably the lattice strain results in smaller fluctuations of the band structure and therewith contribute along with the photogating effect to the faster relaxation exponent of the photoconductivity ($\tau_1 \approx 1$ day, Fig. 1c) , whereas the atomic vacancies cause the deep lying states[21, 22, 31] resulting in the slower relaxation time ($\tau_2 \approx 34$ days, Fig. 1c).

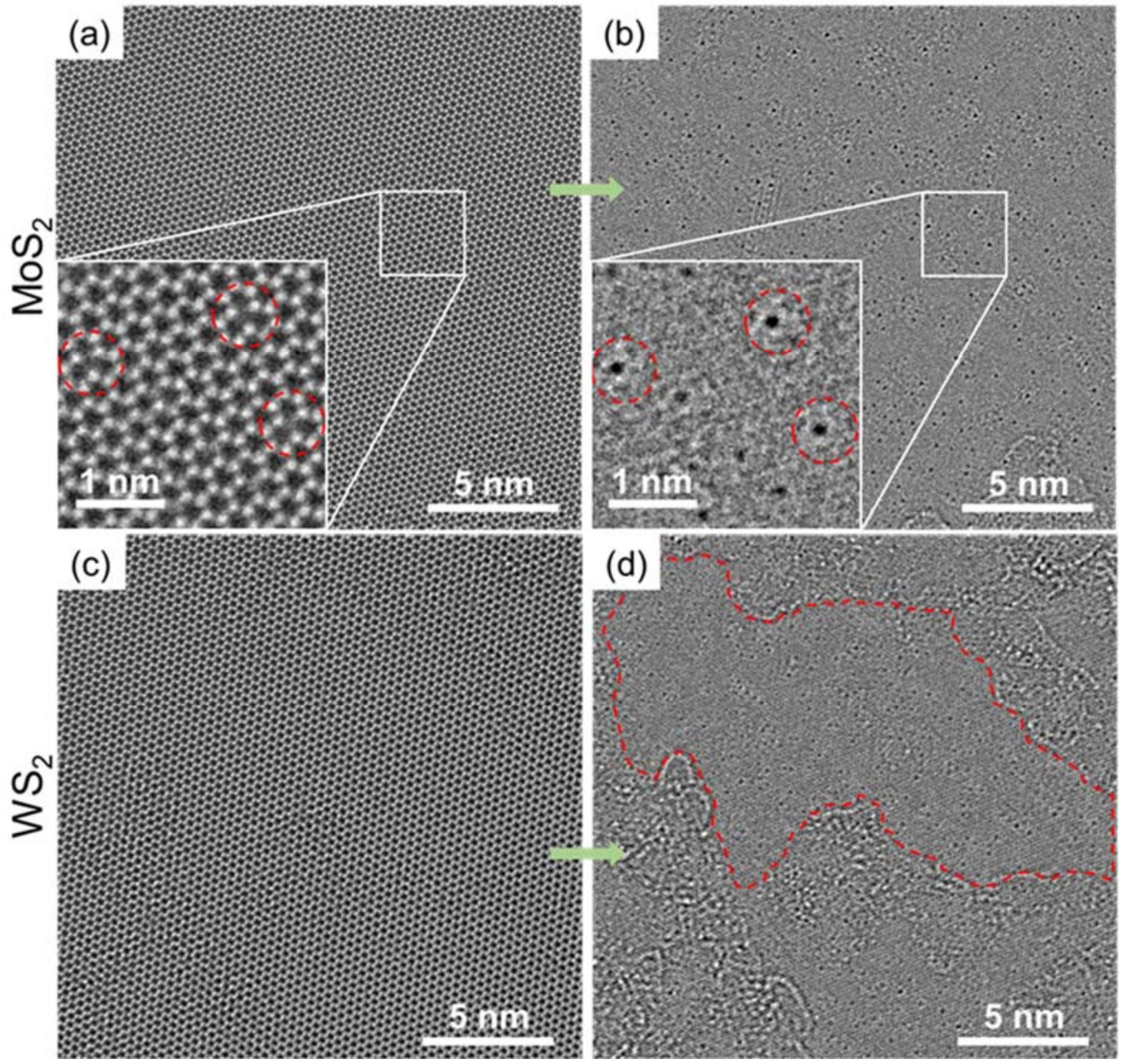

**Figure 4.** 60 kV chromatic (Cc) and spherical (Cs) aberration-corrected HRTEM images of CVD grown ML-$MoS_2$ (a-b) and ML-$WS_2$ (c-d). (a) shows the raw image of the ML-$MoS_2$. The area within the white square is magnified in the lower left. A few red circles mark vacancies, which are difficult to see even in the magnified image. Thus, Fourier-filtering was applied to remove the frequencies of the $MoS_2$ lattice, which is shown in (b). In (b), the same area like in (a) is magnified. Due to the Fourier-filtering the vacancies are better visible (black dots, surrounded by red circles). The same procedure was also applied for $WS_2$ (c) raw image and (d) Fourier-filtered image). Due to the filtering, contaminations become also more visible outside the framed area, thus only clean areas were evaluated with certainty for the defect concentration.

**Comparison with monolayer $WS_2$-FETs**

To further support the defect induced origin of the observed GPPC in $MoS_2$-FETs, we carried out a comparative study of $WS_2$-FETs made of CVD grown monolayers. Evaluation of the HRTEM data presented in Fig. 4c-d shows that in this case the total intrinsic concentration of sulfur vacancies in ML-$WS_2$ is 0.49(9) vac/nm$^2$, which is a factor of 1.6 lower than in ML-$MoS_2$. Additionally, the concentration of $S_2$ vacancies is about 0.022(5) vac/nm$^2$, which means that the relative concentration of $S_2$ vacancies in ML-$WS_2$ in comparison to ML-$MoS_2$ is a factor of 3 lower. In agreement with this evaluation, we found that in $WS_2$-FETs the PPC is significantly weaker in comparison to $MoS_2$-FETs (see Fig. S16). At similar conditions the PCC reveals a time constant ($\tau$) of only ~6 hours. In contrast to $MoS_2$-FETs our control experiments with $WS_2$-FETs show that rather the external factors, i.e. the adsorbates, the monolayer/substrate interaction and photogating[4, 6, 7] than the internal structural defects contribute to the observed PPC of in the latter devices (Fig. S17, SI Part 2 for details).

**Modification of the optical emission by UV irradiation**

Finally, in addition to the modification of the transport properties by UV irradiation, we also expect that the optical properties of the ML-$MoS_2$ are modified. After the irradiation, a significant amount of VB electrons is excited and localized in the trap states below the CB, which has to result in a significantly quenching of the photoluminescence (PL). We performed PL emission mapping of as grown ML-$MoS_2$ crystals on $SiO_2$/Si substrate before and after UV irradiation, see Fig. 5a and Fig.5b, respectively. As can be seen, after the irradiation with UV irradiation the PL emission is significantly diminished, which is in agreement with our expectation. We further tested this effect by preparing suspended ML-$MoS_2$ on TEM grids and observed similar behavior (see SI Fig. S18 and SI Part 6 for details). Thus, the GPPC induced by UV irradiation also allows one to effectively modify the related physical properties of ML-$MoS_2$.

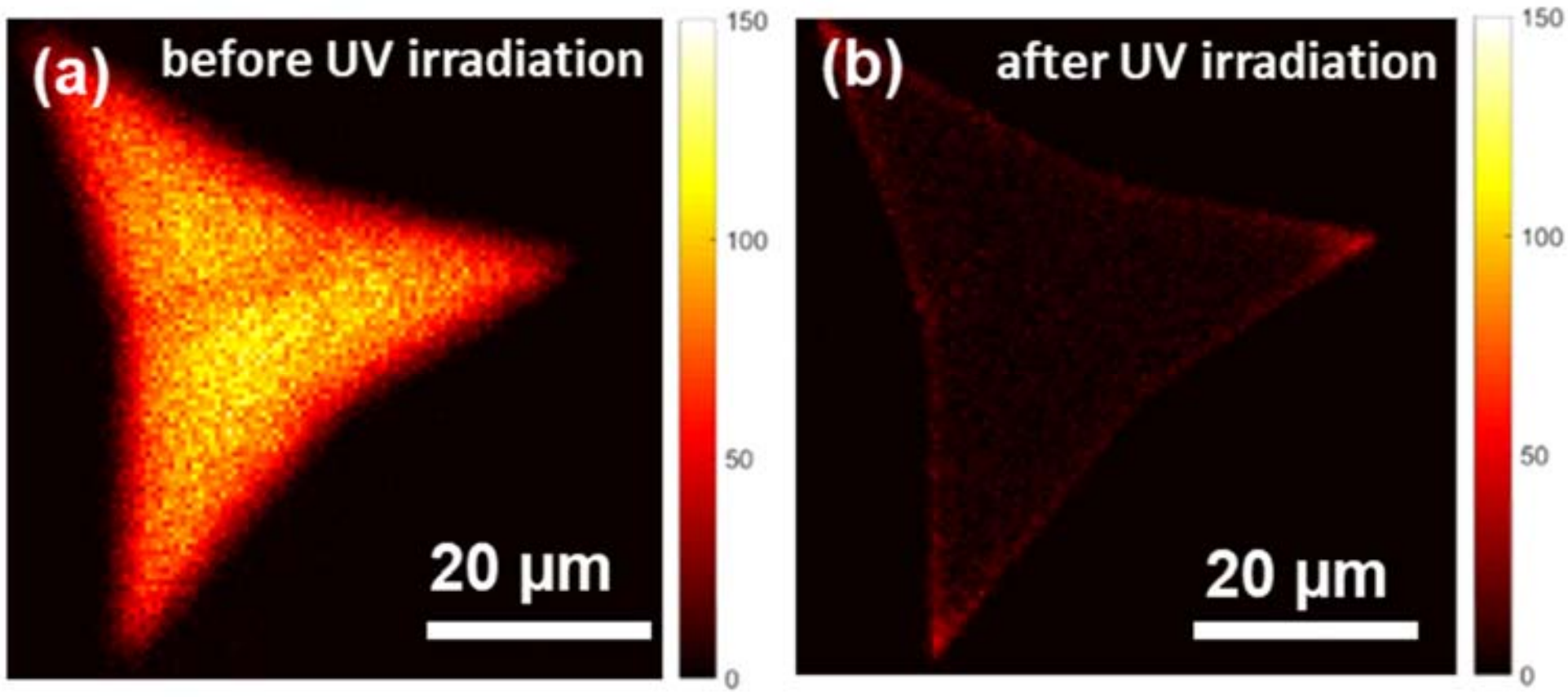

**Figure 5.** (a-b) Photoluminescence (PL) maps of an as-grown ML-$MoS_2$ crystal on a $SiO_2$/Si substrate before (a) and after irradiation (b) with UV light ($\lambda$ = 365 nm, ~30 mW/cm$^2$, 5 min). The PL intensity significantly diminishes after exposure to UV light.

## Conclusions

In summary, we presented an experimental observation of the giant persistent photoconductivity (GPPC) in the CVD grown ML-MoS2 after their irradiation with UV light. The respective time constant of the slow GPPC component is about 30 days at RT. Our theoretical study suggests that the GPPC results from the large spatial fluctuations of the random potential energy of the charge carriers (photo-electrons and photo-holes) in the ML-$MoS_2$, leading to their significant spatial separation and therefore prolonged recombination time. This description is supported by the spectroscopy and microscopy study of the ML-$MoS_2$ on the atomic scale, which allows us to identify the atomic vacancies in the monolayer as a major factor for the observed GPPC effect. Besides the transport properties, we also demonstrate that the related optical properties such as PL emission are significantly diminished in the ML-$MoS_2$ after UV irradiation. Our results shed light to the fact that atomic defects play a crucial role in the optoelectronic properties of ML-TMDs and it is highly essential to understand defect related properties to further develop the field of TMD based electronics and optoelectronics. Furthermore, efficient routes towards defect engineering of ML-TMDs may enhance their applicability. We anticipate that the GPPC effect can be effectively exploited further for applications in electronic, optoelectronic and biotechnological devices (see, e.g., Refs. [32, 33, 34, 35, 36]).

## Methods

The ML-TMDs used in this study were synthesized by CVD method. [10, 11] The basic properties of the monolayers after the synthesis were characterized by Raman and X-ray photoelectron spectroscopy as well as by optical and atomic force microscopy. After this characterization, the FET devices were microfabricated by electron beam lithography. The electric transport measurements before and after the UV irradiation ($\lambda$ = 365 nm) were conducted in a vacuum probe station (~$10^{-6}$ mbar). The PL measurements were conducted at RT using the excitation wavelength of 532 nm. The electronic band structure of the monolayers down to the atomic scale was characterized by low temperature (1.1 K) scanning tunneling spectroscopy measurements. The chromatic (Cc) and spherical (Cs) aberration-corrected high-resolution low electron energy TEM was conducted at RT using the SALVE microscope.[27] All experimental procedures and the data evaluation are presented in SI in details.

## Data availability

Data presented in this study are available on request from the authors.

## Acknowledgements

We acknowledge financial support of the Thüringer MWWDG via the 'ProExzellenz 2014–2019' programme under the grants 'ACP$^{Explore}$2016' and 'ACP$^{Explore}$2018' as well as the Deutsche Forschungsgemeinschaft (DFG) through a research infrastructure grant INST 275/257-1 FUGG and CRC 1375 NOA (Project B2). This project has also received funding from the joint European Union's Horizon 2020 and DFG research and innovation programme FLAG-ERA under grant TU149/9-1. T.L. and U.K. acknowledge funding from the DFG and the Ministry of Science, Research and the Arts (MWK) of the federal state of Baden-Württemberg (Germany) in the frame of the SALVE project (www.salve-project.de) as well as the European Union in the frame of the Graphene Flagship. M.V.F. acknowledges the financial support of the Ministry of Education and Science of the Russian Federation in the framework of Increase Competitiveness Program of NUST "MISiS" K2-2020-001. We thank Stephanie Höppener and Ulrich S. Schubert for enabling our Raman spectroscopy and microscopy studies at the Jena Center for Soft Matter (JCSM).

**Author contributions**

A.G., M.V.F. and A.T. conceived the research and designed the experiments. A.G performed the FET measurements analyzed all data. M.V.F. developed and performed the theoretical analysis. M.G., M.S., and T.F. synthesized the hBN monolayer, performed the STS measurements and analyzed these data with respect to the observed transport behavior. T.L performed the HRTEM measurement and analysis supervised by U.K. R.M. and I.S. performed the PL imaging and analysis. U.H performed the device fabrication. D.K contributed to transport measurements and interpretation of results. C.N performed the AFM and Raman spectroscopy measurements and analysis. A.G., N.M., and A.L.M. synthesized the $MoS_2$ and $WS_2$ monolayer crystals. A.G, M.V.F. and A.T. wrote the manuscript with inputs from all authors.

**Additional information**

Correspondence and request for materials should be addressed to A.T. (andrey.turchanin@uni-jena.de).

**Competing financial interests**

The authors declare no competing interests**.**

# Supporting Information

## Giant persistent photoconductivity in monolayer $MoS_2$ field-effect transistors

A. George[1, 2], M. V. Fistul[3,4,5], M. Grünewald[6], D. Kaiser[1], T. Lehnert[7], R. Mupparapu[2,8], C. Neumann[1], U. Hübner[9], M. Schaal[6], N. Masurkar[10], A. L. M. Reddy[10], I. Staude[2,8], U. Kaiser[7], T. Fritz[6], A. Turchanin[1, 2*]

*[1]Friedrich Schiller University Jena, Institute of Physical Chemistry, 07743 Jena, Germany*
*[2]Abbe Centre of Photonics, 07743 Jena, Germany*
*[3]Institute for Basic Science (IBS), Center for Theoretical Physics of Complex Systems, 34126 Daejeon, Republic of Korea*
*[4]Ruhr-University Bochum, Theoretische Physik III, 44801 Bochum, Germany*
*[5]National University of Science and Technology (MISiS), 119049 Moscow, Russia*
*[6]Friedrich Schiller University Jena, Institute of Solid State Physics, 07743 Jena, Germany*
*[7]Ulm University, Central Facility of Electron Microscopy, Electron Microscopy Group of Materials Science, 89081 Ulm, Germany*
*[8]Friedrich Schiller University Jena, Institute of Applied Physics, 07745 Jena, Germany*
*[9]Leibniz Institute of Photonic Technology, 07745 Jena, Germany*
*[10]Wayne State University, Department of Mechanical Engineering, 48202 Detroit, USA*

e-mail: andrey.turchanin@uni-jena.de
Tel.:+49-3641-948370
Fax: +49-3641-948302

## Table of contents

## 1. Experimental

### *1.1 CVD growth of monolayer $MoS_2$ (ML-$MoS_2$) and monolayer $WS_2$ (ML-$WS_2$)*

ML-$MoS_2$ and ML-$WS_2$ crystals were grown by the CVD process.[1, 2] Silicon substrates with a thermally grown $SiO_2$ layer of 300 nm were used as substrates (Siltronix, roughness 0.3 nm RMS). The growth was carried out in a two-zone tube furnace with a tube diameter of 55 mm. The substrates were cleaned initially by ultrasonication in acetone for 5 min followed by washing in isopropanol and blown dry with argon. A quartz crucible containing sulfur powder (99.98%, Sigma Aldrich) was placed in the center of the first zone of the tube furnace. The substrates were placed next to a wafer containing ~1 µg $MoO_3$ powder (99.97%, Sigma Aldrich) for $MoS_2$ growth or 5 mg $WO_3$ (99.99%, Alfa Aesar) mixed with 250 µg NaCl for $WS_2$ growth and loaded to the center of the second zone of the furnace. Subsequently, the quartz tube was evacuated to $5 \times 10^{-2}$ mbar pressure and refilled with argon. The growth was carried out at atmospheric pressure under an argon flow of 100 $cm^3$/min. The second zone containing the metal oxide precursor and the substrates were heated to the growth temperature of 770 °C (for $MoS_2$) or 860 °C (for $WS_2$) at a rate of 40 °C/min and held at that temperature for 15 min. The sulfur temperature was adjusted to reach 200 °C when the second zone reaches 750 °C (for $MoS_2$) or 800 °C (for $WS_2$). A flow of $H_2$ gas at a rate of 10 $cm^3$/min was introduced to the chamber during the growth time. After the growth, the furnace was turned off and allowed to cool down under an argon flow of 100 $cm^3$/min until 350 °C were reached. Then the tube furnace was opened to rapidly cool down to room temperature (RT). These procedures result in the growth of ML-$MoS_2$ and ML-$WS_2$ crystals of mainly triangular shape with a typical size of about 50-100 µm.

### *1.2 Preparation of the hexagonal boron nitride (h-BN) monolayer on Pt(111)*

The Pt(111) single crystal (purchased from MaTecK GmbH) was cleaned by means of several $Ar^+$ sputtering (1 kV, 4 µA, $5\times10^{-5}$ mbar, 30 min) and annealing (800°C for 30 min) cycles. The absence of contaminations was controlled by X-ray photoelectron spectroscopy (XPS). For XPS an x-ray source (SPECS XR50) with Al Kα excitation (1486.7 eV) in combination with a hemispheric electron analyzer (SPECS EA-200) was used.

For the growth of the 2D h-BN layer we followed the steps described by Orlando et al.[3] They showed that the adsorption of borazine on Ir(111) at room temperature followed by an annealing step at 800°C reduces the number of different adsorption configurations and results in an h-BN layer exhibiting a low defect density. We thus exposed the substrate to borazine ($B_3N_3H_6$, Katchem spol. s.r.o.) vapor with a pressure of $5x10^{-8}$ mbar for 10 min with the substrate being held at room temperature. Immediately after that we increased the substrate temperature to 800°C for 10 min followed by an annealing step at 800°C for another 10 min without borazine exposure in order to guarantee the complete dehydrogenation of the precursor molecules on the substrate. This growth progress is known to be monolayer-terminated.[3, 4]

The quality of the h-BN layer was verified with XPS. Fig S11b shows the corresponding B 1s and N 1s spectra. We used a polar angle of 70° to increase the surface sensitivity of the measurement. Both spectra show only one component, which was modelled by an asymmetric Mahan line shape.[5] The asymmetry of the peak can be explained by the slight corrugation of the h-BN layer as well as by scattering of the photoelectrons at electronic states near the fermi edge.[5] The amount of contaminations like oxygen and carbon was less than 1%. Point defects in the h-BN layer should cause a second component in the B 1s and N 1s spectra at lower binding energy which is not visible in the observed XP spectra[6]. In conclusion, our results show a high-quality h-BN layer which is a suitable substrate for STS investigations of a $MoS_2$ film.

*1.3 Basic characterization by optical microscopy, atomic force microscopy and Raman spectroscopy of the grown monolayers*

After the CVD growth the transition metal dichalcogenides (TMDs) were characterized by optical microscopy (OM), atomic force microscopy (AFM) and Raman spectroscopy (see Figs. S2, S3). The OM images were taken with a Zeiss Axio Imager Z1.m microscope equipped with a 5 mega pixel CCD camera (AxioCam ICc5) in bright field operation.

The AFM measurements were performed with an Ntegra (NT-MDT) system in contact mode at ambient conditions using n-doped silicon cantilevers (CSG01, NT-MDT) with a typical tip radius of 6 nm and a typical force constant of 0.03 $Nm^{-1}$.

The Raman spectra were acquired using a Bruker Senterra spectrometer operated in backscattering mode. Measurements at 532 nm were obtained with a frequency-doubled

Nd:YAG Laser, a 50× objective and a thermoelectrically cooled CCD detector. The spectral resolution of the system is 2–3 $cm^{-1}$. For all spectra the Si peak at 520.7 $cm^{-1}$ was used for peak shift calibration of the instrument. The Raman spectrum shown in Fig. S2b reveals the characteristic peaks of ML-$MoS_2$ at 384$cm^{-1}$ and 404 $cm^{-1}$, which are originated from the in-plane ($E^1_{2g}$ band) vibrations of the Mo-S bonds and out-of-plane ($A_{1g}$ band) vibrations of S atoms in the $MoS_2$ lattice,[7] respectively. The difference between the peak positions is 20 $cm^{-1}$ confirming that the crystal is a monolayer.[7]

*1.4 Preparation of ML-$MoS_2$ and ML-$WS_2$ FET devices*

After the growth, the ML-$MoS_2$ and ML-$WS_2$ crystals were transferred onto the device substrates (Siltronix, heavily p-doped silicon substrates with thermally grown $SiO_2$ layer of 300 nm)[2]. To transfer, a PMMA layer of 200 nm (950 kDa, All-Resist, AR-P 679.04) was spin coated onto the $SiO_2$ substrate with CVD grown crystals and hardened for 10 min at 90 °C. Then the substrate was kept floating on top of a bath of KOH solution to etch away the $SiO_2$ layer and to release the monolayer crystals supported by PMMA, followed by washing several times with ultrapure water (18.2 MΩcm, Membrapure) to remove any residual KOH. Then the PMMA supported crystals were placed on the marked $SiO_2$/Si chips and baked at 90 °C for 10 min, followed by immersion in acetone for 2 h to remove the PMMA support. For defining the source and drain electrodes we employed e-beam lithography (EBL). A PMMA resist layer was spin coated on top of the samples, patterned by EBL (Vistec EBPG 5000plus) and subsequently developed. Then the Au/Ti (30 nm/ 5 nm) electrodes were deposited by e-beam evaporation process followed by the dissolution of the PMMA resist in acetone for 2h. The heavily p-doped silicon base functioned as the gate electrode and 300 nm thermal oxide functioned as the gate dielectric.

*1.5 Electrical transport measurements.*

The electrical characterization was carried out with two Keithley 2634B source measure units (SMU). One SMU was used to change the voltage of the gate ($V_g$) with respect to the source/drain in the range between -60 and 60 V for the back-gated devices in vacuum (~$10^{-6}$ mbar). The other SMU was used to apply the source-drain voltage ($V_{ds}$) and measure the source-drain current ($I_{ds}$). A Lakeshore cryogenic vacuum needle probe station TTPX was used to measure the devices in vacuum at a residual pressure about $10^{-6}$ mbar. The mobility is extracted from the linear region of the transfer curve using the equation $\mu = \left(\frac{dI_{ds}}{dV_{bg}}\right)\left(\frac{L}{WC_{ox}V_{ds}}\right)$, where *L* is the channel length, *W* is the channel width, $C_{ox}$ is the capacitance of the 300 nm gate oxide and $V_{ds}$ is the source-drain voltage[2]. For UV irradiation a light emitting diode (LED) with a wavelength of 365 nm (Thorlabs, M365L2), with a typical power output of 360 mW, was used. Also LEDs with wavelengths 455 nm (Thorlabs, M455L3) and 617 nm (Thorlabs, M617L4), respectively, were used to irradiate the devices. The LEDs were controlled using a Thorlabs four-channel LED driver (DC4100).

*1.6 Scanning tunneling microscopy and spectroscopy (STM/STS)*

CVD grown ML-$MoS_2$ crystals were transferred onto a hBN layer grown on a single crystalline Pt(111) substrate. The hBN monolayer thereby serves as an atomically flat, electrically insulating overlayer. The quality of hBN was confirmed by means of scanning tunneling microscopy (STM) (JT-STM/AFM from SPECS Surface Nano Analysis GmbH operated at 1.1 K with tungsten tips) (Fig. S11) before transferring the $MoS_2$ crystals. Prior to the STM/STS experiments, the sample was thoroughly degassed in ultrahigh vacuum at about 120 °C for 2h. For STS we directly measured the derivative of the tunneling current dI/dV using the lock-in technique. We recorded STS spectra in a grid with dimensions of 50 nm × 50 nm and 1 nm spacing in each direction (2500 spectra in total). No hysteresis was observed between forward (1.5 V → -2.75 V) and backward (reverse) bias sweeps, which proves that we do not permanently influence the band structure by the measurement process itself. In order to visualize the spatial distribution of trap states we plotted their onsets as color maps as shown in Fig 3b and 3c.

*1.7 High resolution transmission electron microscopy (HRTEM)*

The HRTEM images were acquired with the Cc/Cs-corrected Sub-Angstrom Low-Voltage Electron microscope (SALVE)[8]. A voltage of 60 kV was used with typical dose rates of about $10^5$ $e^-/nm^2s$. The values for the chromatic aberration Cc and the spherical aberration Cs were between -10 µm to -20 µm. All Cc/Cs-corrected HRTEM images were acquired with bright atom contrast and recorded on a 4k x 4k camera with exposure times of 1 s.

*1.8 Photoluminescence (PL) measurements*

PL from ML-$MoS_2$ crystals was characterized with a MicroTime 200 laser-scanning confocal fluorescence microscope from PicoQuant GmbH. A pulsed laser of wavelength 532 nm and repetition rate of 80 Hz was used to excite the ML-$MoS_2$ crystals and measure their PL emission with a single-photon avalanche diode (SPAD) detector. A microscope objective of 40x magnification and a numerical aperture 0.65 was used to focus the laser onto the crystals, forming a spot of diameter ~1 μm. The PL emission was collected using the same objective. PL maps were acquired by raster scanning of the microscope objective and collecting the PL emission in the spectral range of 650 - 720 nm using a band pass filter, essentially to collect the A-exciton and trion emissions. Care was taken in all the measurements to block the excitation light reaching the detector using a dichroic mirror and a notch filter for the excitation wavelength of 532 nm, in addition to band pass and long pass filters.

## 2. Influence of substrates and adsorbates on the GPPC

To further investigate the reason of the GPPC and to answer the question whether it is of extrinsic or intrinsic origin, we performed electrical measurements as proposed by Kis et al.[9] It is suggested that the surrounding material of the $MoS_2$ channel, especially the substrate, plays an important role in photocurrent dynamics and the PPC by providing traps for charge carriers, which, however, can be discharged by applying a short gate pulse.[9] In Fig. S7, we show the rise of $I_{ds}$ by irradiation with UV light. After the removal of the UV irradiation, the $I_{ds}$ remains nearly constant. We then applied several short gate pulses from -60 V to 0 V to test whether the transfer curves can be restored to the initial state. The effect of gate pulses is minimal as can be seen in Fig. S7. The GPPC behaviour was also tested after the application of the gate pulses and it can be seen that the effect is still present and the device shows similar GPPC decay behaviour even after the application of the gate pulses (Fig. S8).

Furthermore, we have studied the hysteresis of the $MoS_2$-FETs. The respective transfer curves between forward and backward $V_g$ sweeps of UV irradiated $MoS_2$-FET devices are shown in Fig. S9, S10. If a significant density of charge traps was existing at the interface between the $MoS_2$ channel and the $SiO_2$ surface, one would expect significant hysteresis in the transfer characteristics of the FETs.[10, 11, 12] Absence of large hysteresis indicates that additional charge traps either do not exist in significant concentration in our $MoS_2$-FET samples, or do not contribute much to the here observed GPPC. Also, it is observed that the hysteresis of the devices remained the same after the application of several gate pulses (Fig. S10).

We conducted the gate pulse experiments on the $WS_2$-FET devices as well where only a PPC effect is present instead of GPPC (Fig S16). In Fig. S17, the rise and fall of $I_{ds}$ by irradiation with UV light is shown. In this case, the $I_{ds}$ values before UV irradiation can be restored by applying a single short gate pulse as shown in Fig. S17. Thus, we cannot exclude a substrate effect to contribute to the PPC in our samples. Since the ML-$WS_2$ samples have fewer defects (as revealed in our HRTEM study) in comparison with ML-$MoS_2$, the substrate effect is more prominent. However, for the case of $MoS_2$-FETs, the substrate seems to play a minor role, and the defect related spatial fluctuations in the band structure is the prominent mechanism of GPPC.

**3. Evaluation of the defect density by HRTEM**

For the error evaluation, the evolution of the defect concentration was analysed for $MoS_2$ and $WS_2$ (cf. fig. 4). Furthermore, a linear behaviour for low defect concentrations (displaced S-atoms < 5 %) is assumed. Based on the error of the slope, *ΔS*, for the linear defect evolution,

$$\Delta S = \sqrt{\left(\frac{\Delta V}{A \cdot \phi}\right)^2 + \left(\frac{\mathrm{V} \cdot \Delta \mathrm{A}}{A^2 \cdot \phi}\right)^2 + \left(\frac{\mathrm{V} \cdot \Delta \phi}{A \cdot \phi^2}\right)^2},$$

the error of the intrinsic defect concentration, $\Delta C$, is determined using the intersection point at time $t$=0 with the error margin due (red dashed lines) to the calculated slopes of the linear fit as shown in the example for one data set of $WS_2$ in Fig. S14. In total, three data sets for $WS_2$ and also for $MoS_2$ were evaluated. For the confidence intervals, we took $\Delta V = \sqrt{V}$ for the vacancies and $\Delta A = \sqrt{A}$ for the evaluated area. For the total accumulated dose, a huge confidence interval is assumed because of the uncertainties of the previous electron beam irradiation before the first image was recorded. Here, we assumed an error for the irradiation time of $t$ = 10s, so that the confidence interval for the accumulated dose, depending on the dose rate $\varphi$, becomes $\Delta\phi = 10s \cdot \varphi$ (Fig. S14).

In Fourier filtered Cc/Cs-corrected HRTEM images the location of the vacancies can be clearly identified but not the defect type. Therefore, the differentiation between double and single sulphur vacancies was performed in the unfiltered image by checking the contrast of each vacancy. Fig. S15 shows a Cc/Cs-corrected HRTEM image of $MoS_2$ with single ($S_1$) and double ($S_2$) sulfur vacancies. For differentiation, the intensity profile along the line-scans are given (blue for double $S_2$ vacancy and red for the single $S_1$ vacancy) showing that the contrast for single $S_1$ vacancies is noticeable reduced compared to a pristine S column. Due to the absence of the sulfur atoms in the $S_2$ vacancy, no S-peak occurs in the profile (blue curve).

**4. Estimation of the lattice strain in CVD grown ML-$MoS_2$**

Typically, ML-$MoS_2$ grown by CVD at high temperature conditions are subjected to significant tensile strain due to the mismatch of thermal expansion coefficients of monolayers and the $SiO_2$/Si substrate.[13] Furthermore, these monolayers are also subjected to electron doping from the $SiO_2$ substrate. The strain inside a CVD grown monolayer is estimated using a well-known analysis based on spectral positions of first-order Raman modes as described previously.[14, 15] Employing the Grüneisen parameters ($\gamma$), electron doping factors ($k_n$) from the literature,[16, 17] the strain is estimated relative to the exfoliated ML-$MoS_2$ , which are typically considered to be free of strain. The strain in ML-$MoS_2$ grown with the CVD method is estimated to be 0.34±0.08 %. The parameters considered in the estimation of strain inside the ML-$MoS_2$ are shown in the table below.

**Table 1:** Parameters used for estimation of the lattice strain

| | | |
|---|---|---|
| Positions of first-order Raman modes for CVD grown ML-$MoS_2$ (in $cm^{-1}$) | 383.9±0.4 ($E^1_{2g}$) | 404.7±0.4 ($A_{1g}$) |
| Positions of first-order Raman modes for exfoliated ML-$MoS_2$ (from [3]) (in $cm^{-1}$) | 385.6 ($E^1_{2g}$) | 405.3 ($A1_g$) |
| Grüneisen parameters ($\gamma$) | $\gamma_{E^1_{2g}}$ | 0.68 |
| | $\gamma_{A_{1g}}$ | 0.21 |
| Electron doping factor ($cm^{-1}/10^{13}\ cm^{-2}\ e^{-1}$) | $k^E_n$ | -0.33 |
| | $k^A_n$ | -2.22 |
| Strain (ε) (in %) | 0.34±0.08 % | |

**5. Effect of visible light irradiation on the transport in $MoS_2$-FETs**

In addition to UV irradiation (365 nm), we also tested irradiation of $MoS_2$-FETs with 455 nm and 617 nm light (Figs. S19, S20), and the devices did not show any GPPC. The devices returned nearly to their initial conditions after several hours. This indicates that irradiation with higher energy photons are required for causing the GPPC, while the exact reason for this is presently not well understood.

**6. Photoluminescence (PL) measurements of suspended ML-$MoS_2$**

By preparing suspended ML-$MoS_2$ samples one could ultimately exclude substrate effects. We have prepared such samples by transferring ML-$MoS_2$ on to Quantifoil type TEM grids. Such a grid has a thin carbon film with an array of holes with a diameter of 2 µm separated by 2 µm. Thus, the ML-$MoS_2$ regions spread over the holes can be considered as (partially) suspended areas. The emission maps of a $MoS_2$ crystal on such a TEM grid before and after UV irradiations (Fig. S18a, b) show a similar behavior as observed in as-grown samples. The areas that appear as bright dots are the suspended regions of the ML-$MoS_2$, and it can be observed that the PL intensity is diminished after the UV irradiation, however, quantitatively not by the same magnitude as observed on $SiO_2$/Si substrates. In order to interpret the UV induced quenching quantitatively, histograms of the PL intensity were extracted from these maps as shown in Fig. S18c, d. The fraction of frequencies of bins with counts in the range of 0 - 40 (the range where PL is absent) increased significantly from 81% to 92%, and bins with counts higher than 40 (the range where PL is present) decreased after the UV exposure. This observation infers that as a result of the UV exposure, the PL signal across the suspended UV irradiated $MoS_2$ crystal is quenched and thus causing a reshaping of the histogram. UV irradiation triggers photoexcitation and charging of traps, which is the prerequisite for the PPC, and thereby suggests that this effect is present even on suspended samples.

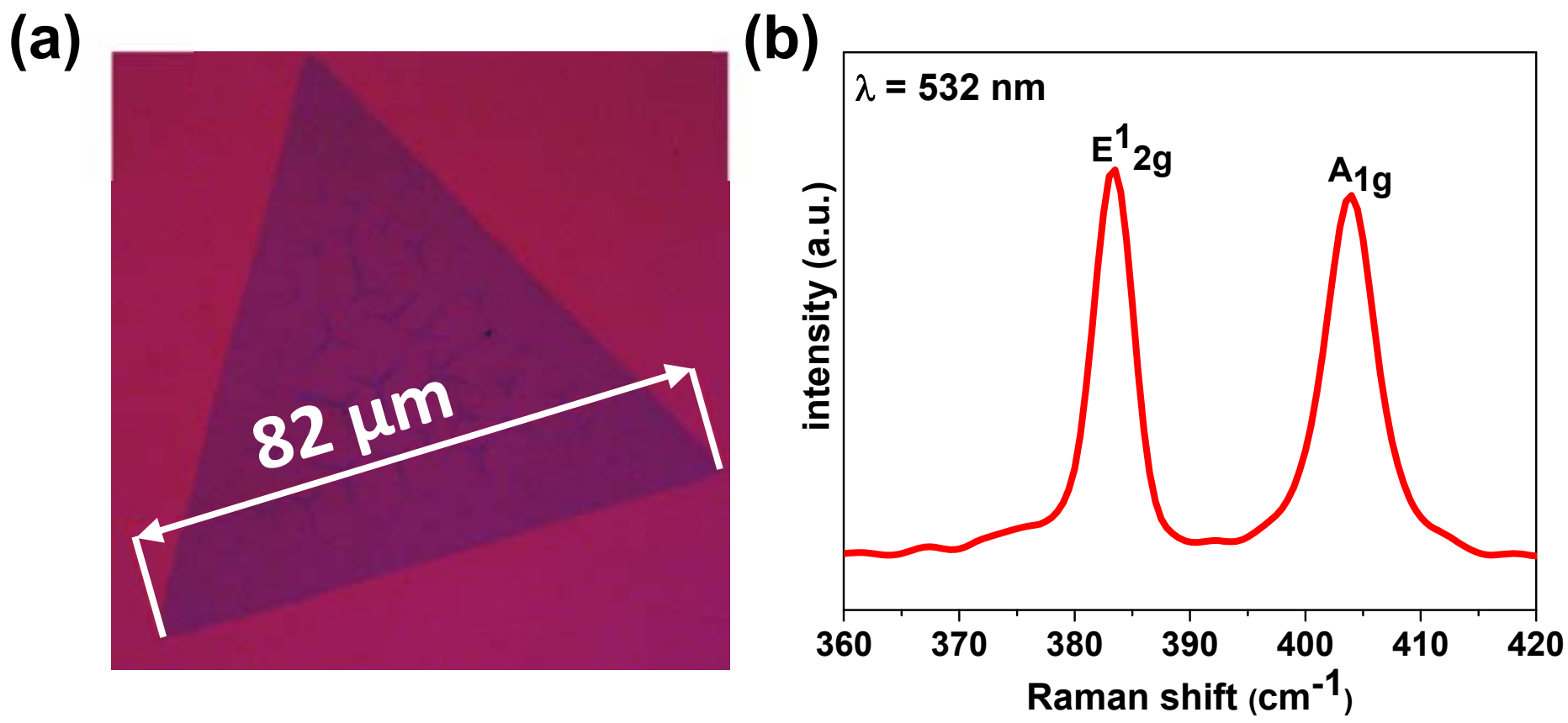

**Fig. S1:** (a) Optical microscopy image of the CVD grown ML-$MoS_2$ on a $SiO_2$/Si substrate. (b) Raman spectrum recorded on a ML-$MoS_2$ crystal. The difference of peak positions is 20 $cm^{-1}$ corresponding to a monolayer.

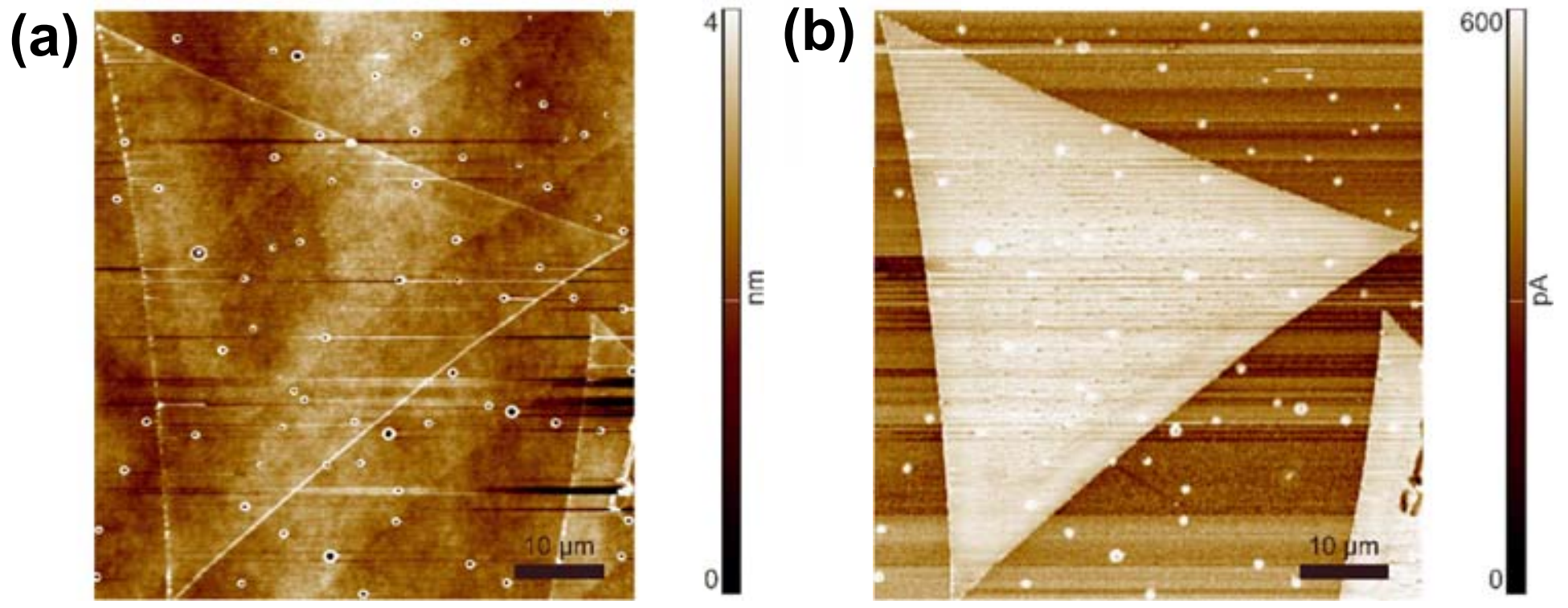

**Fig. S2:** AFM height (a) and lateral force (b) images of a ML-$MoS_2$ crystal.

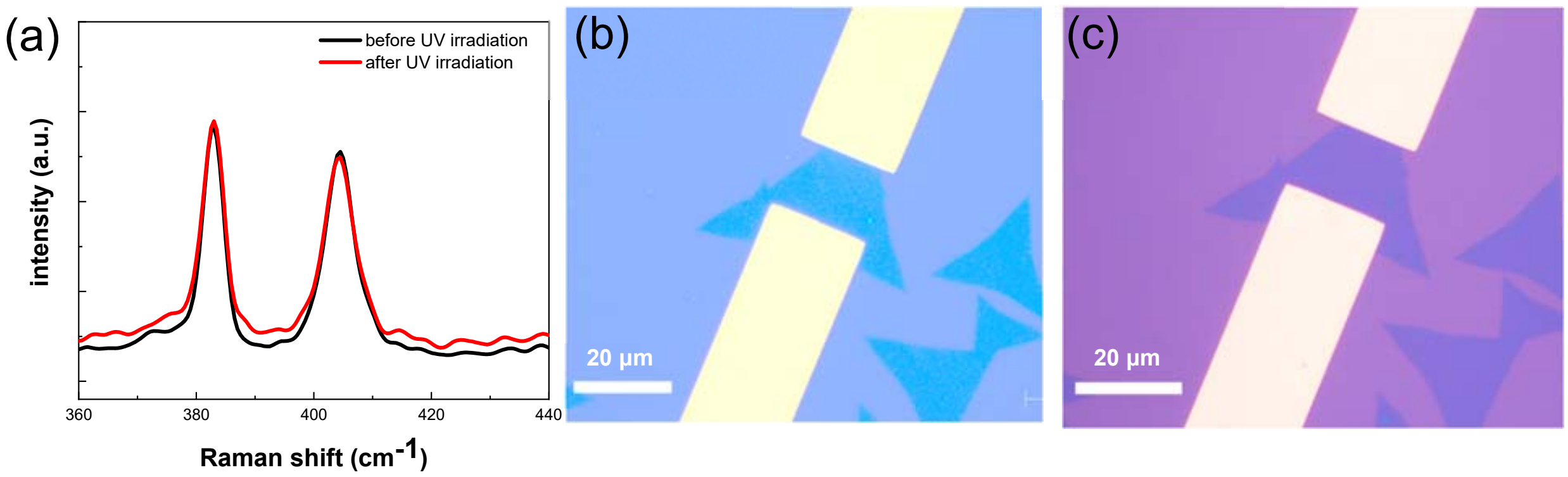

**Fig. S3: (a)** Raman spectra of ML-$MoS_2$ crystal before and after UV irradiation (~30 mW/cm$^2$, 5 min) showing no significant beam induced structural modification or damage. Respective optical microscopy images of the same device before (b) and after (c) UV irradiation. False colors due to different illumination conditions and contrast enhancement.

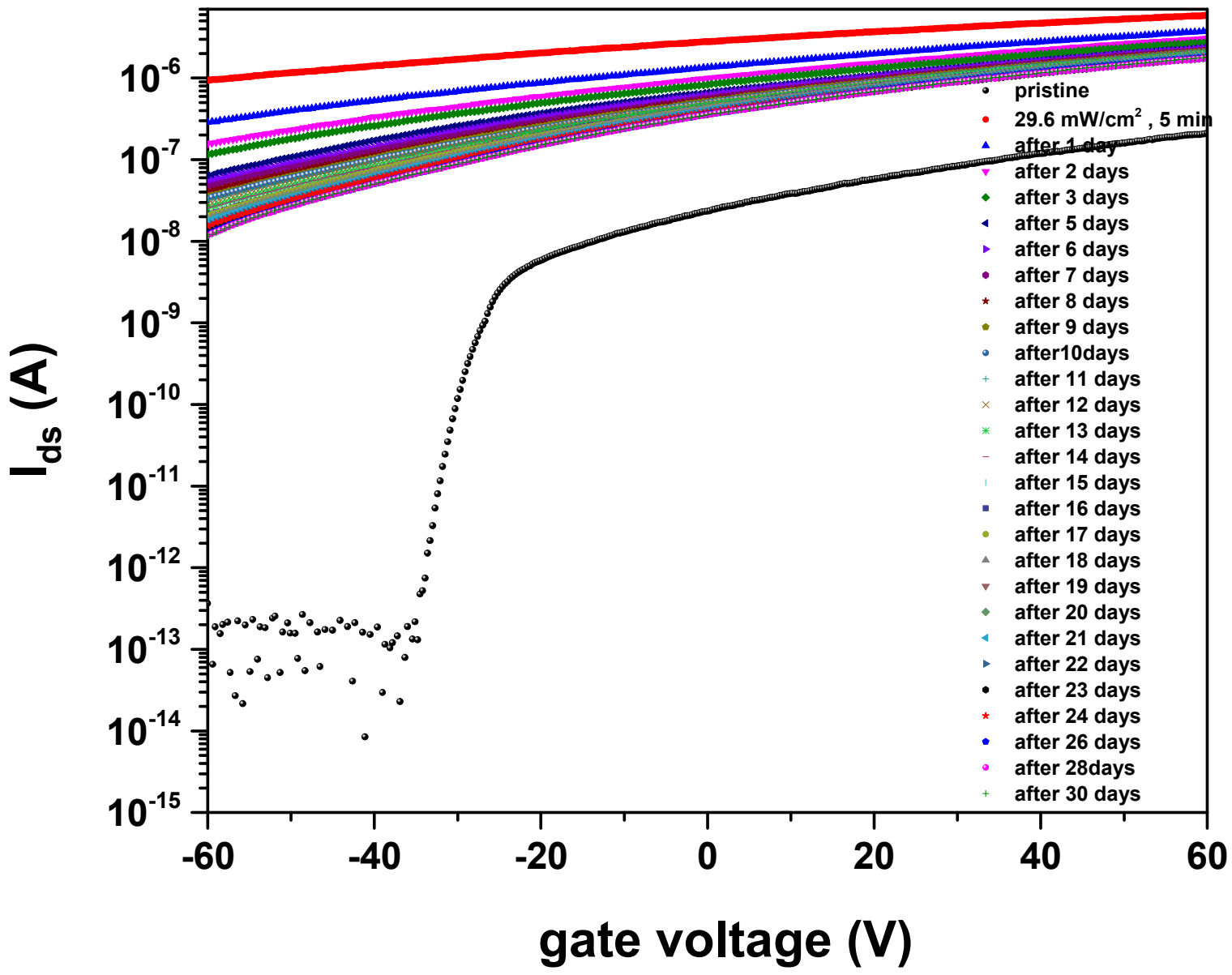

**Fig. S4:** Complete data set of the experimental observation of the GPPC in a $MoS_2$-FET device is shown in the graph. The black curve represents the transfer characteristics of $MoS_2$-FET before UV irradiation, the red curve represents the transfer characteristics immediately after UV irradiation (λ = 365 nm) for 5 min with an intensity of ~30 mW/cm$^2$. The colored curves represent the transfer characteristics recorded after the following days of UV irradiation. The measurements continued up to 30 days.

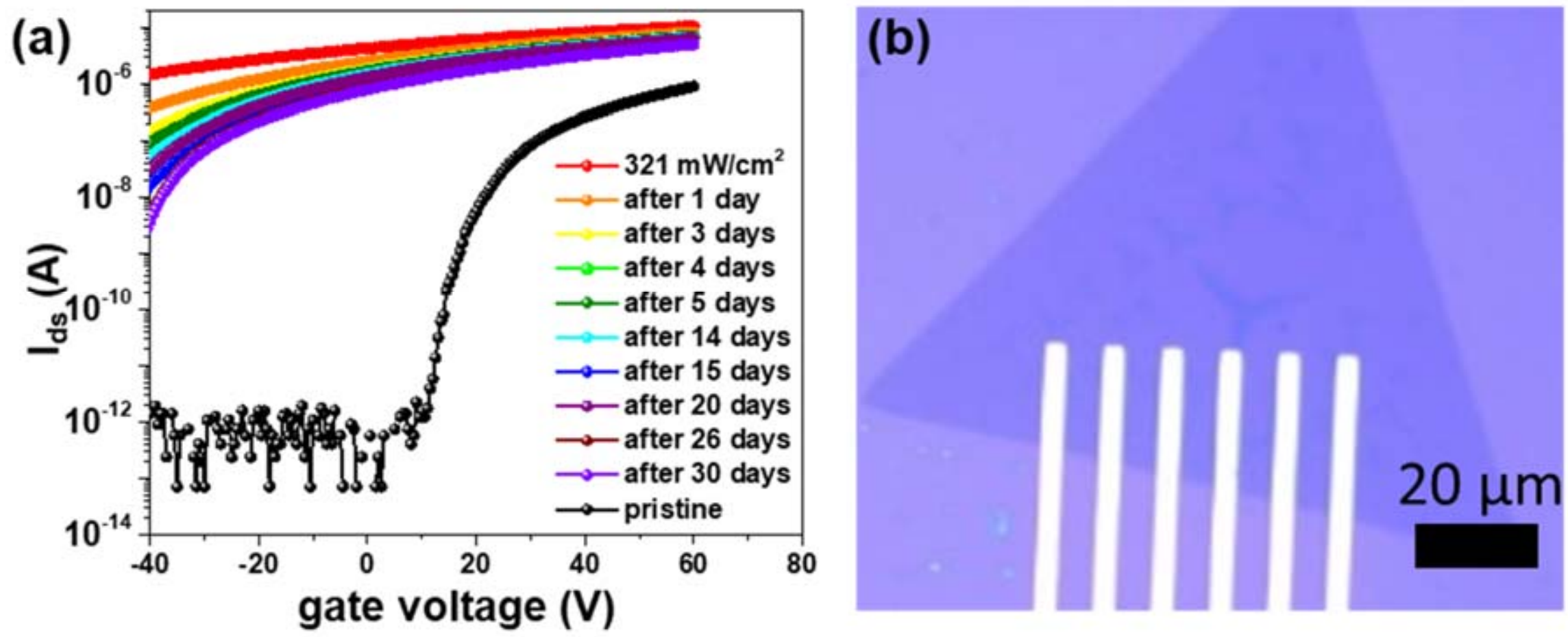

**Fig. S5** (a) GPPC behavior observed in the $MoS_2$-FET device used for the RT and LT measurement is shown in the graph. The black curve represents the transfer characteristics of $MoS_2$-FET before UV irradiation, the red curve represents the transfer characteristics immediately after UV irradiation (λ = 365 nm) for 5 min with an intensity of ~30 mW/cm$^2$. The colored curves represent the transfer characteristics recorded after the following days of UV irradiation. The measurements continued up to 30 days. (b) Optical microscopy image of the $MoS_2$-FET devices used for RT and LT measurements.

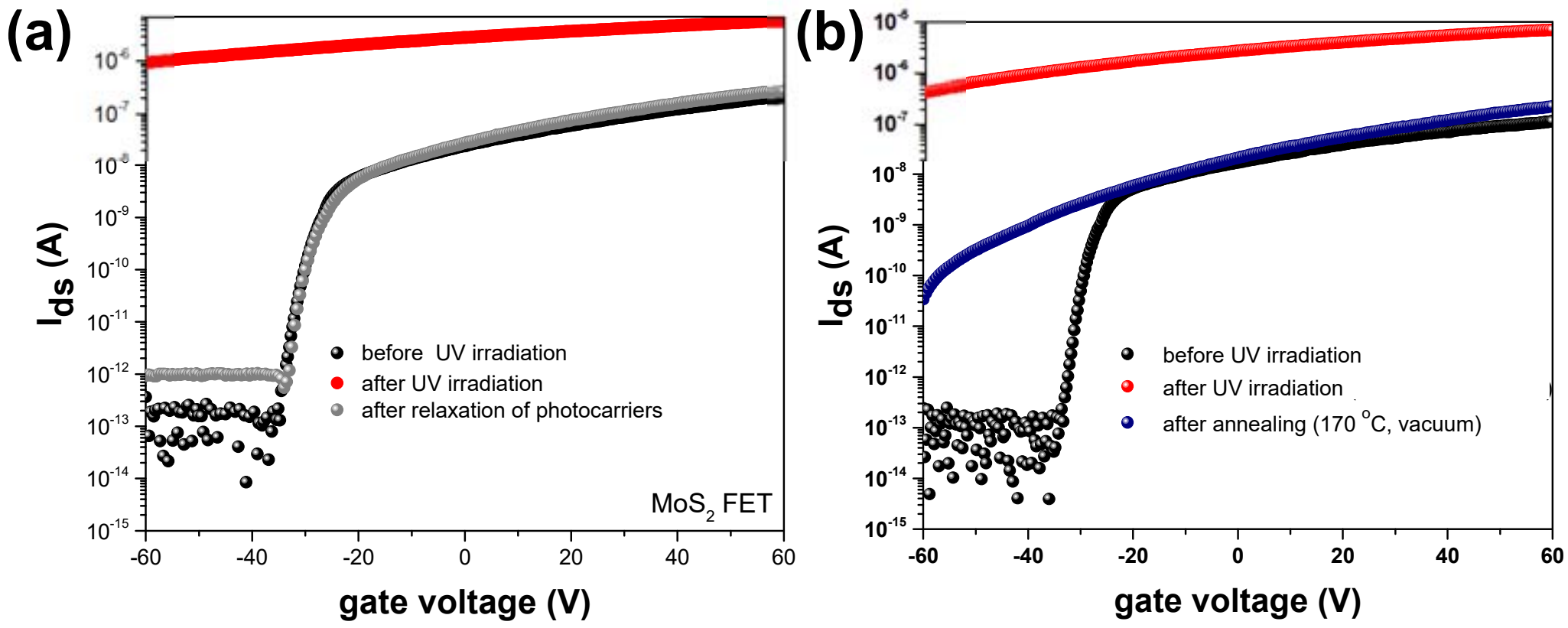

**Fig. S6:** a) After several months (stored in the probe-station under high vacuum), the UV irradiated (~30 mW/cm$^2$, 5 min) $MoS_2$-FET returned to its initial characteristics by complete recombination of photo-generated carriers. b) Furthermore, after annealing at 170 $^{\circ}$C for 3 h in vacuum the device nearly restored to its pristine transfer characteristics after irradiation.

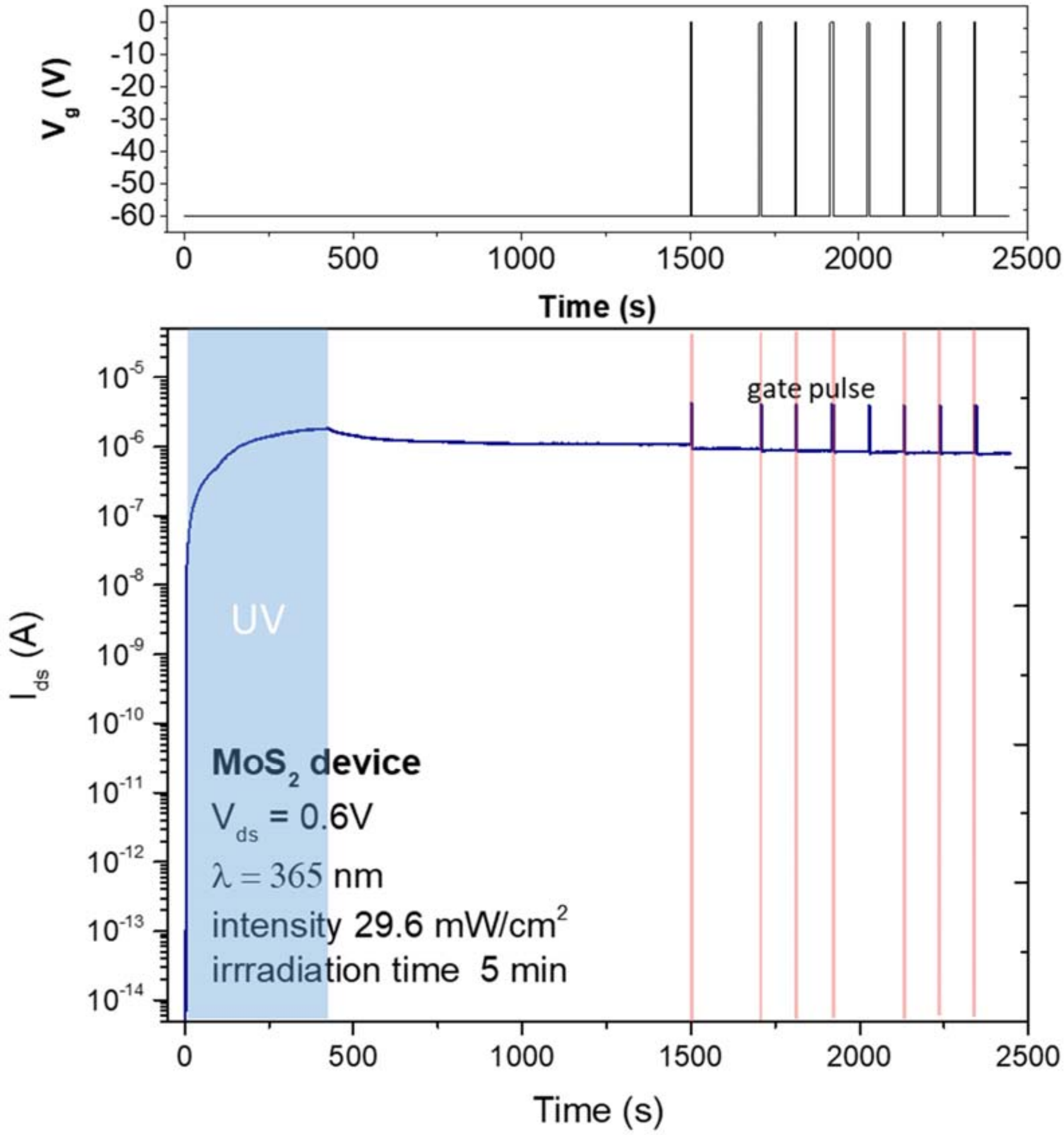

**Fig. S7**. Photocurrent dynamics measurements on a $MoS_2$-FET: Rise and fall of the photocurrent with UV irradiation (λ = 365 nm) is shown. Several short gate pulses from -60 V to 0 V were applied to test the possibility of eliminating any charges trapped at the substrate-$MoS_2$ interface as suggested by Kis et al.[9] It is found that the effect of short gate pulses is minimal.

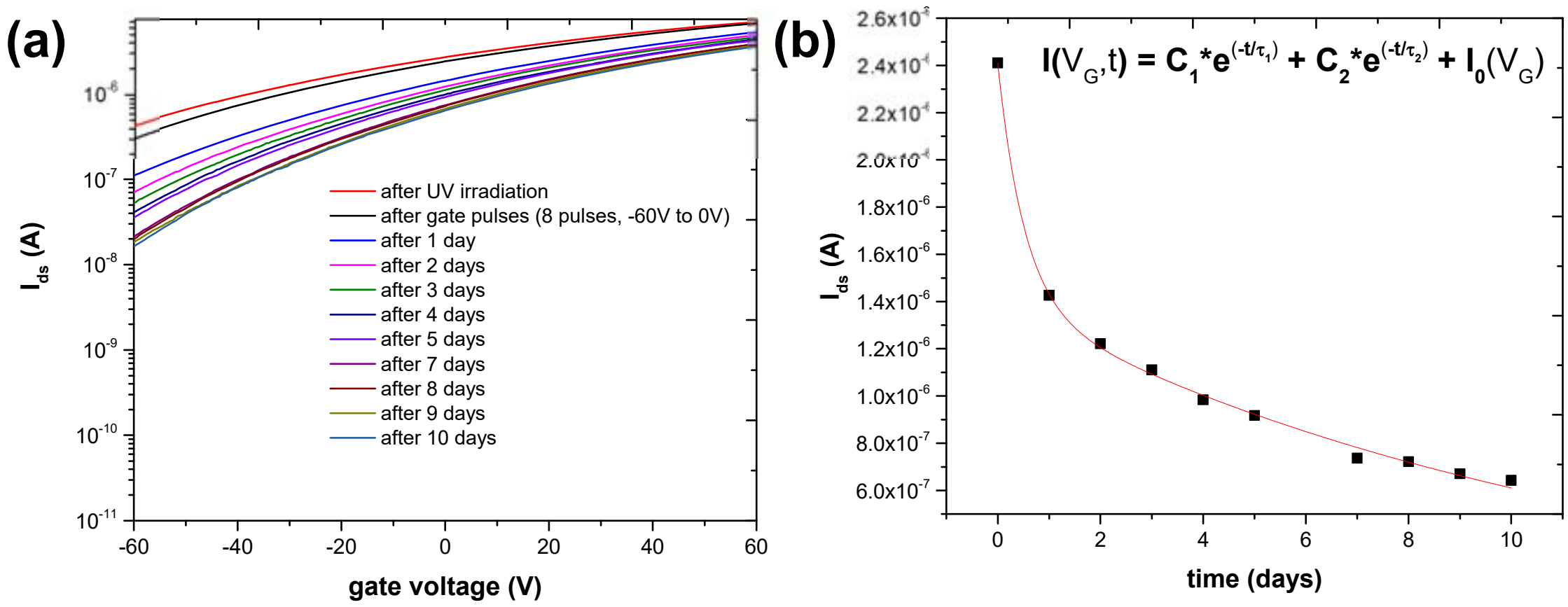

**Fig. S8**. (a-b) Decay of GPPC of a $MoS_2$-FET with time after the application of gate pulses. The decay shows similar behavior with or without the application of short gate pulses.

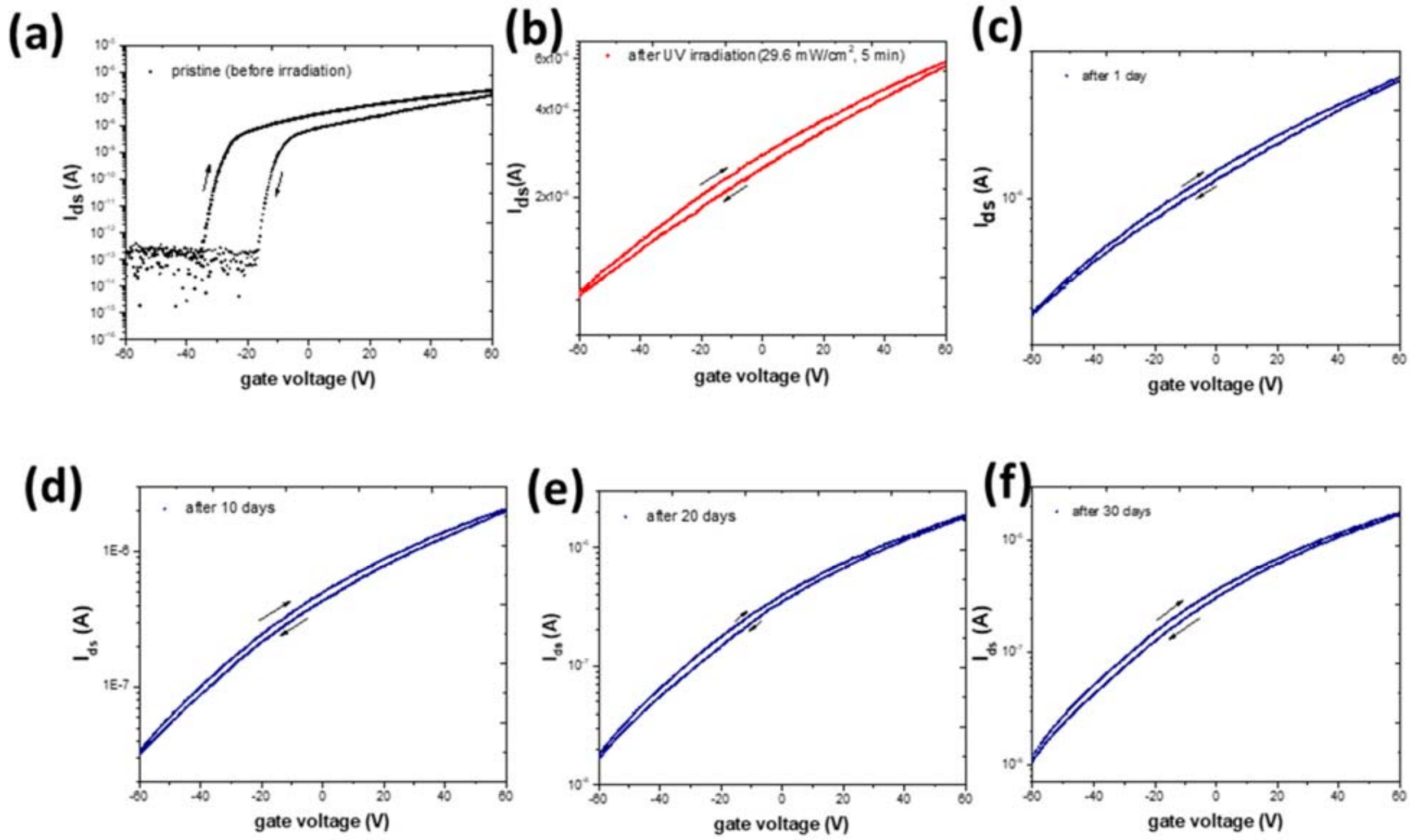

**Fig. S9:** Hysteresis in $MoS_2$-FET before (a) and after UV irradiation (b-f).

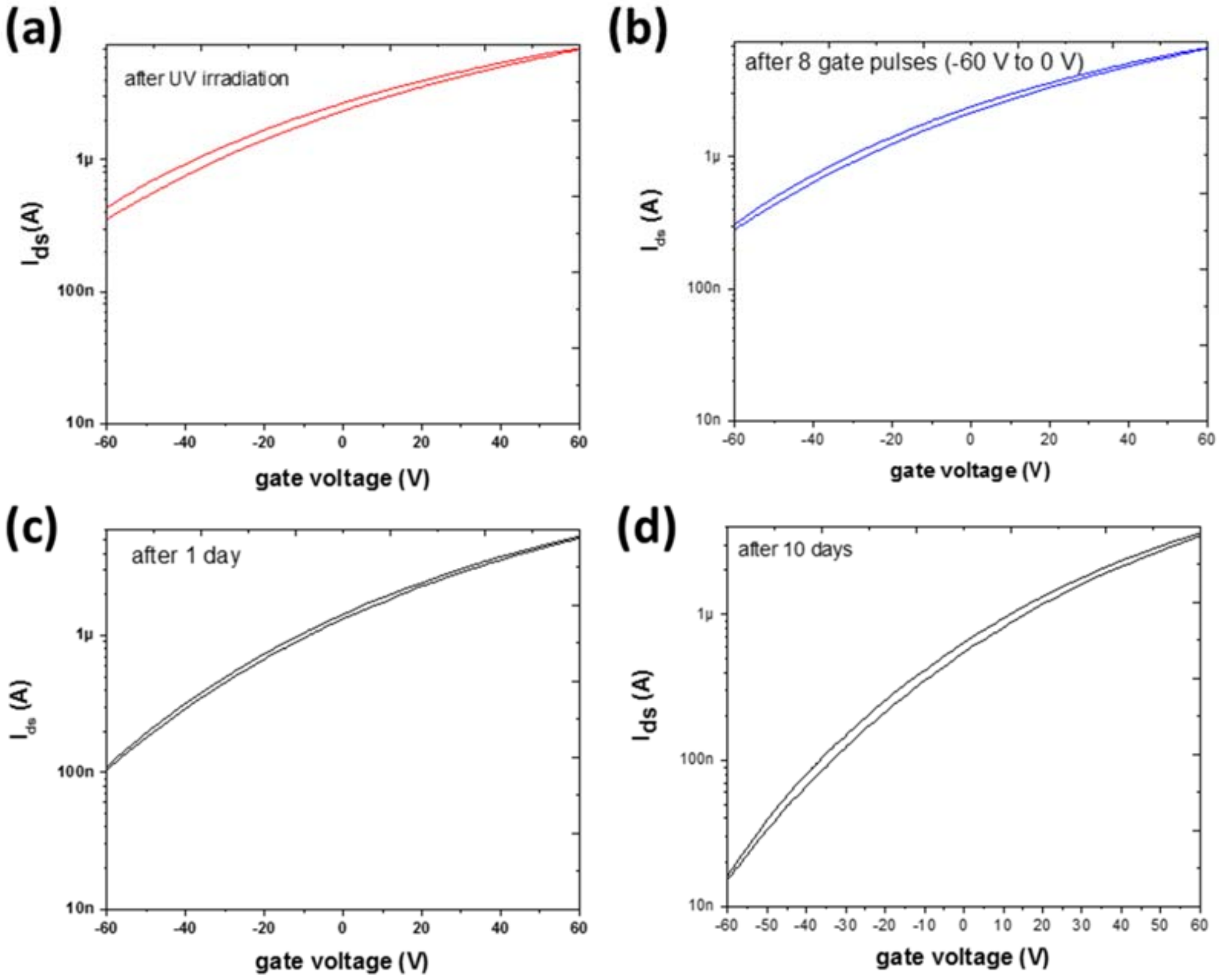

**Fig. S10:** Hysteresis in $MoS_2$-FET just after UV irradiation (a) and after the application of 8 short gate pulses from -60 V to 0 V (b-d).

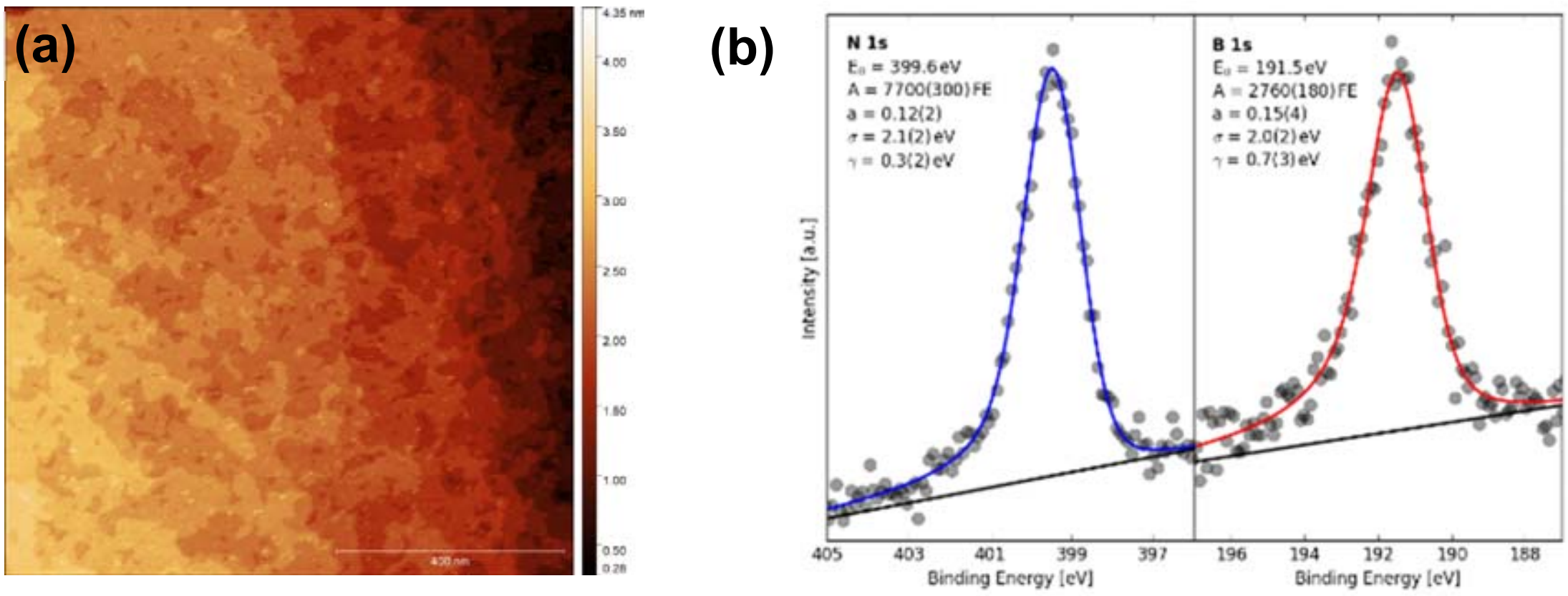

**Fig. S11:** a) STM image (1 µm × 1 µm, $V_T$ = 0.2 V, $I_T = 50pA$) of a pristine monolayer hBN/Pt(111) showing monoatomic steps and large terraces. b) Core level spectra of the pristine h-BN monolayer on Pt (111). Both spectra were acquired at 70° emission angle using an excitation energy of 1486.6 eV (Al $K_\alpha$). Measured data points are given as dots; the solid lines represent the best fits to the data using a Mahan line shape[5]. Backgrounds are represented by linear functions (black lines).

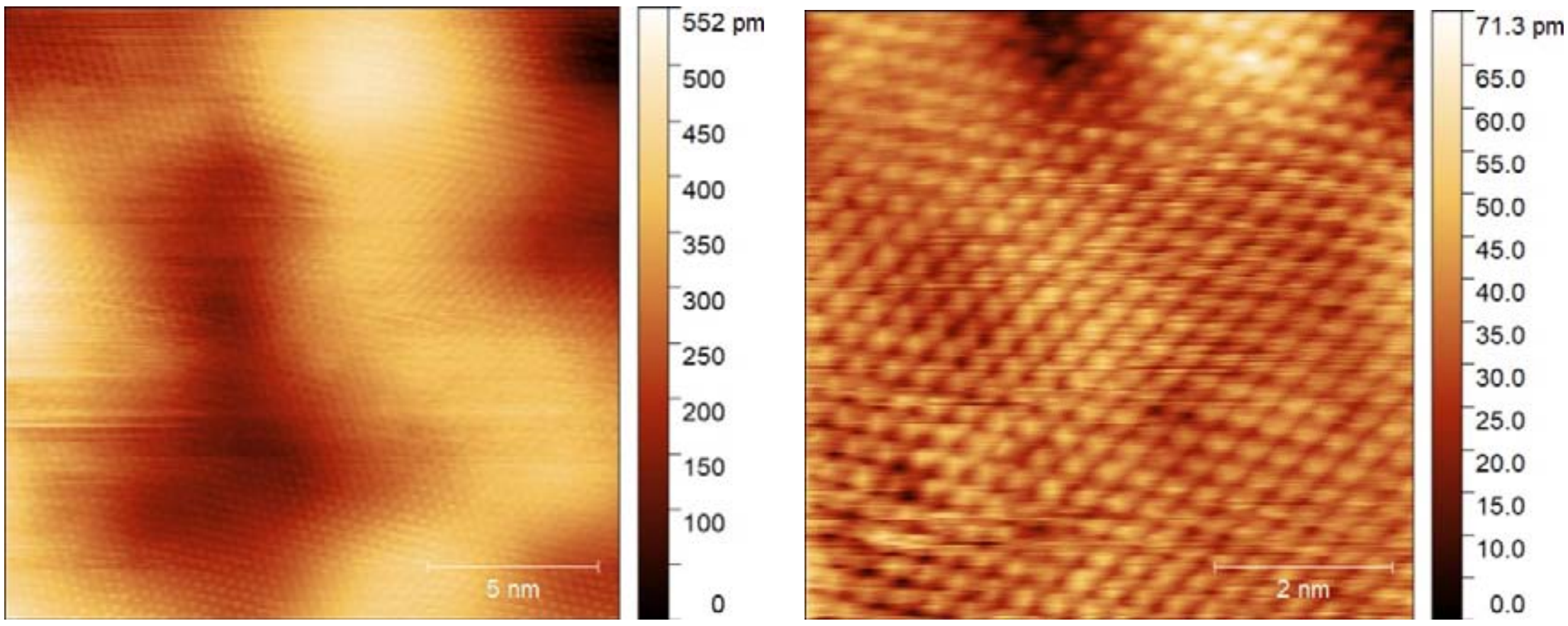

**Fig. S12:** (a) STM image (18 nm x 18 nm, $V_T$ = 1.5 V, $I_T$ = 10 pA) with atomic resolution of ML-$MoS_2$ and (b) separately measured close-up view.

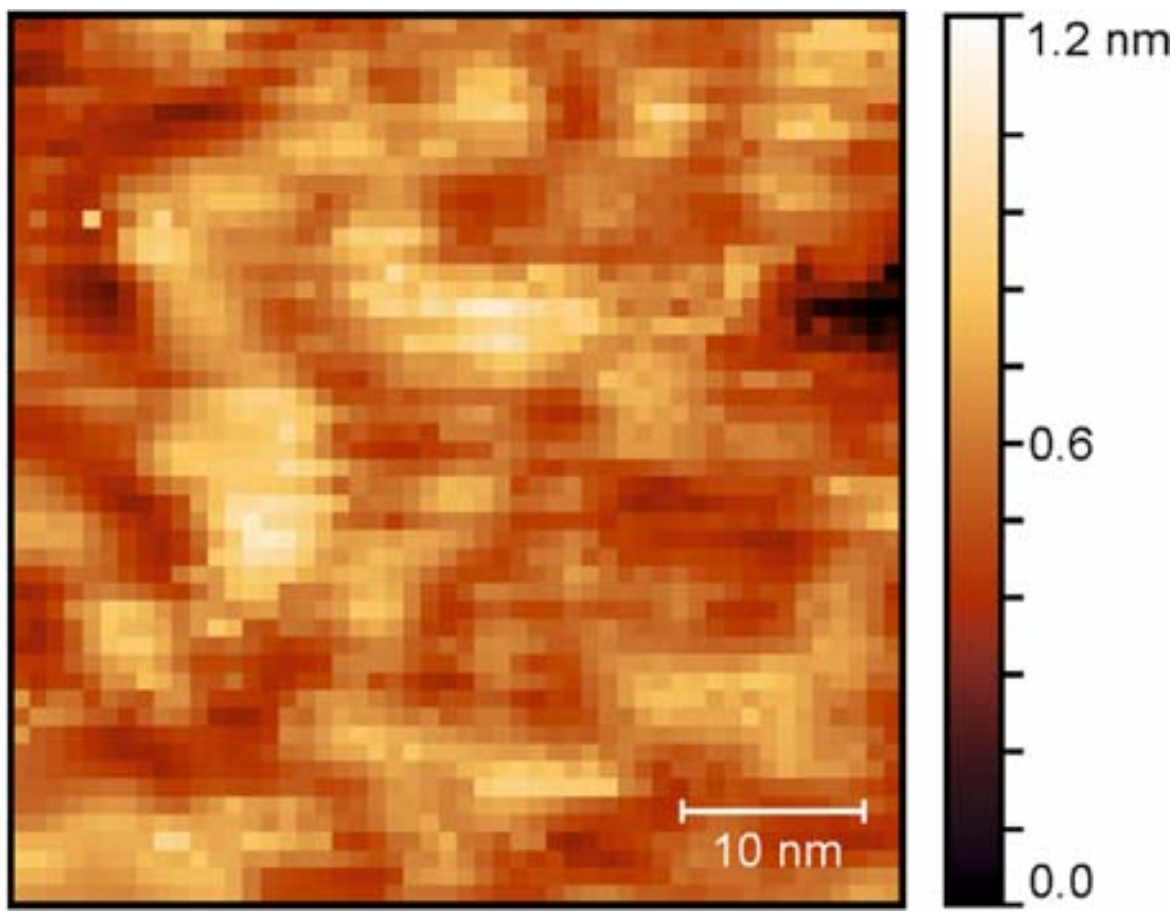

**Fig. S13:** Topography map (50 nm x 50 nm, $V_T$ = 1.5 V, $I_T$ = 500 pA) of ML-$MoS_2$/hBN/Pt(111) as measured together with the STS maps in Fig. 3.

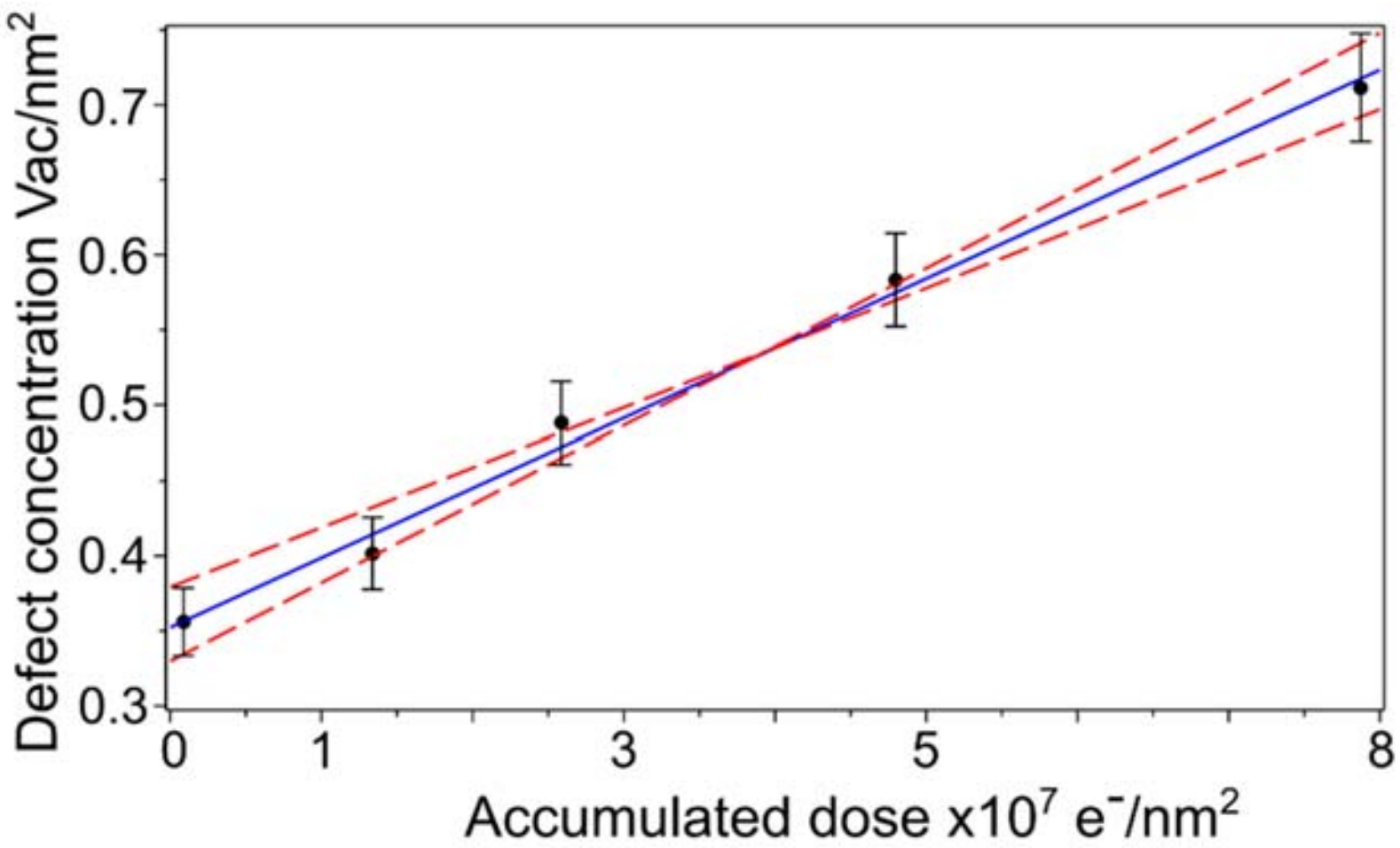

**Fig. S14:** Example for sulfur vacancy concentration determination: Measurements (black dots) of the sulfur vacancies evolution (both $S_1$ and $S_2$) per area depending on the accumulated dose electron dose in $WS_2$. Based on the slope of the linear fit (blue line), the error calculation was carried out. The red dashed lines show the maximum and minimum slopes within the range of the calculated errors. The intrinsic vacancy concentration is determined by extrapolating the number of intrinsic defects to zero electron irradiation dose.

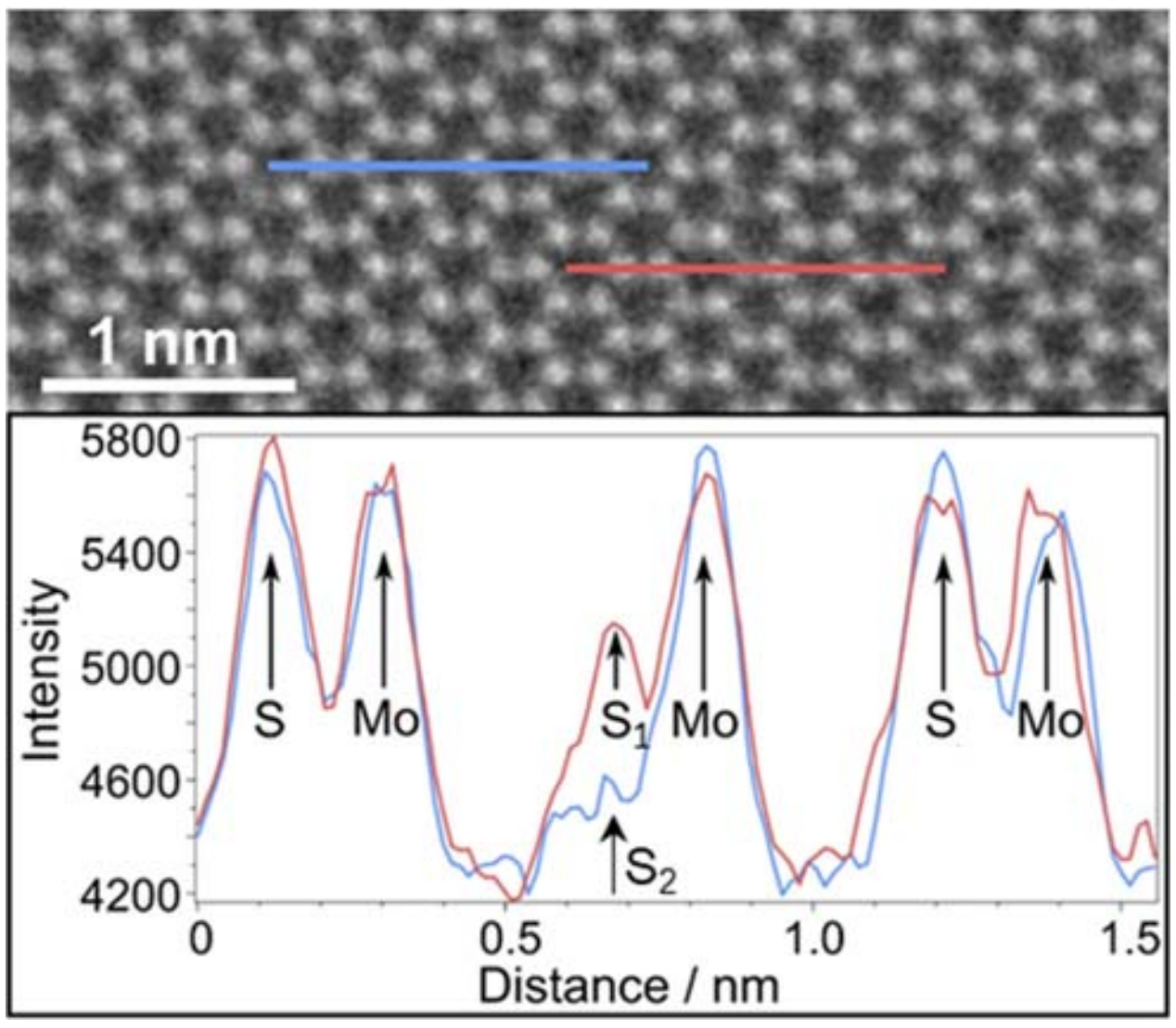

**Fig. S15:** 60 kV Cc/Cs-corrected HRTEM image of ML-$MoS_2$ with single and double vacancies. Blue and red lines mark lines with a double ($S_2$) and a single ($S_1$) vacancy, respectively. The red intensity profile shows that in the middle only one sulfur atom is absent resulting in a contrast reduction by about a half compared to the neighbouring position, where two sulfur atoms are present (marked as S). The blue intensity profile shows in the middle a further decrease of the intensity at the sulfur position due to the absence of sulfur, i.e. a $S_2$ vacancy.

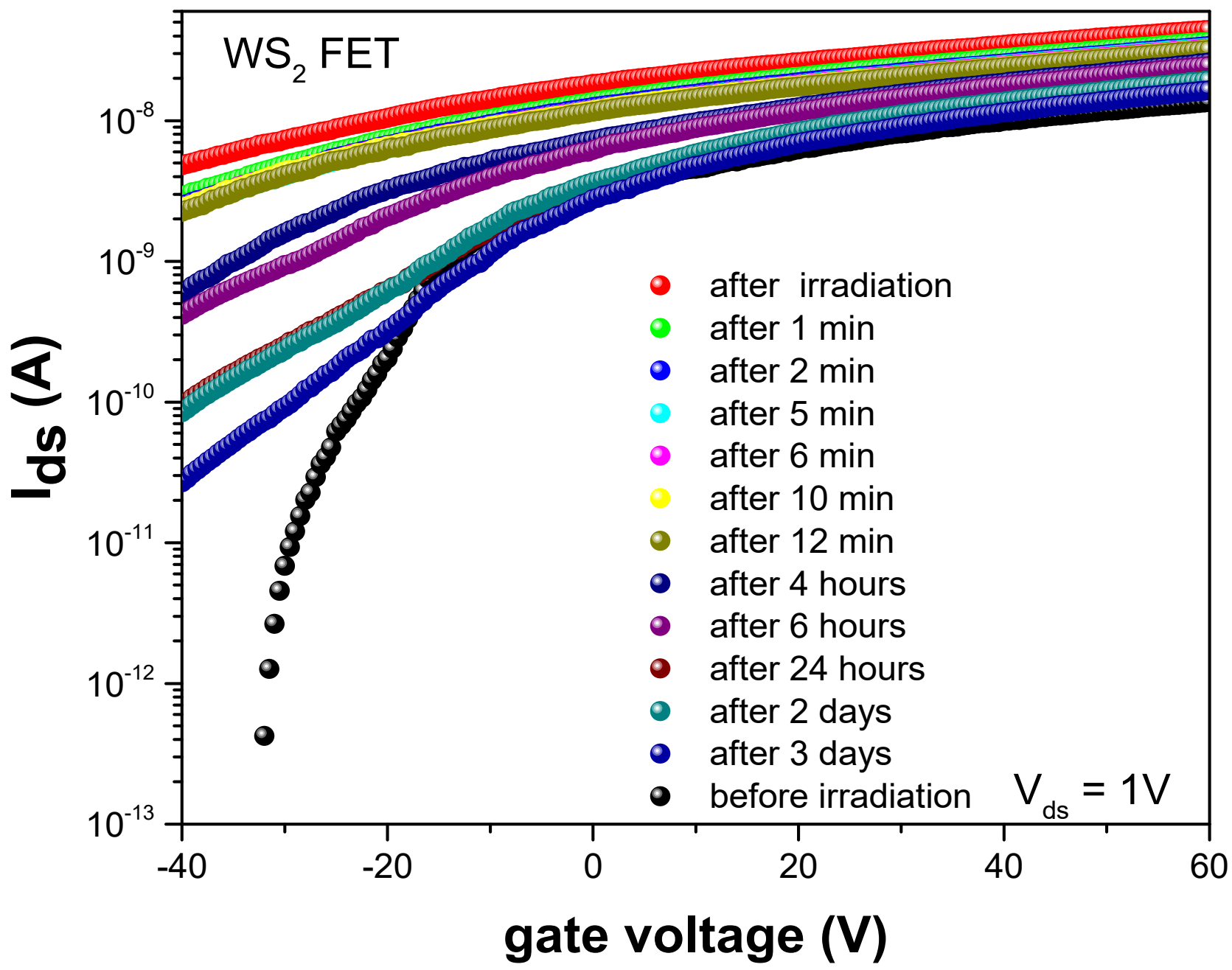

**Fig. S16:** Decay of the PPC effect in a $WS_2$-FET device. Transfer characteristics of a $WS_2$-FET device before (black) and immediately after UV (λ = 365 nm) irradiation (red). The irradiation time was 5 min at an intensity of ~30 mW/cm$^2$. The colored curves indicate the decay of the PPC with time.

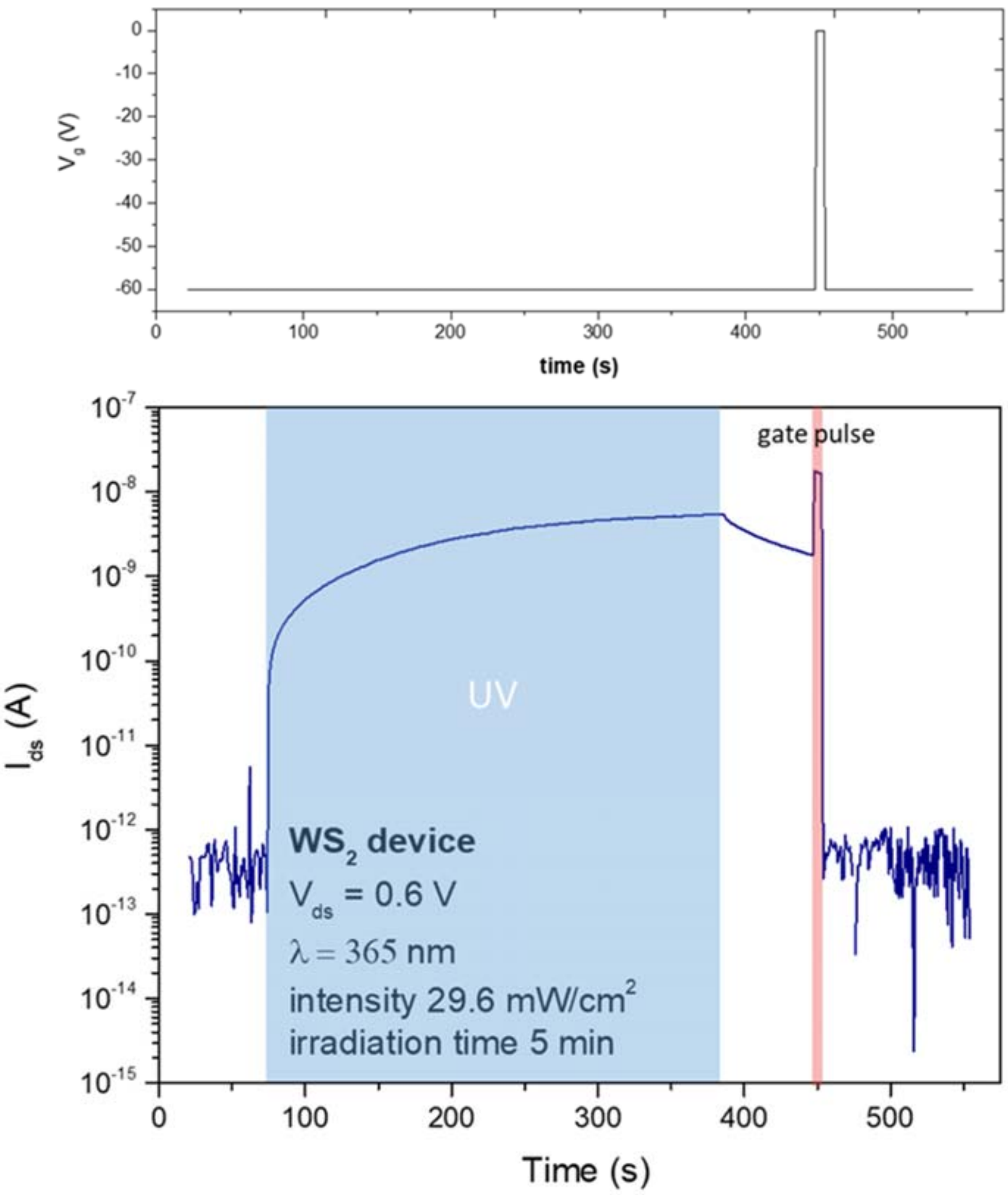

**Fig. S17:** Photocurrent dynamics measurements on $WS_2$-FET: Rise and fall of the photocurrent with UV irradiation and after application of gate pulses from -60 V to 0 V the drain current restored to the initial value.[9]

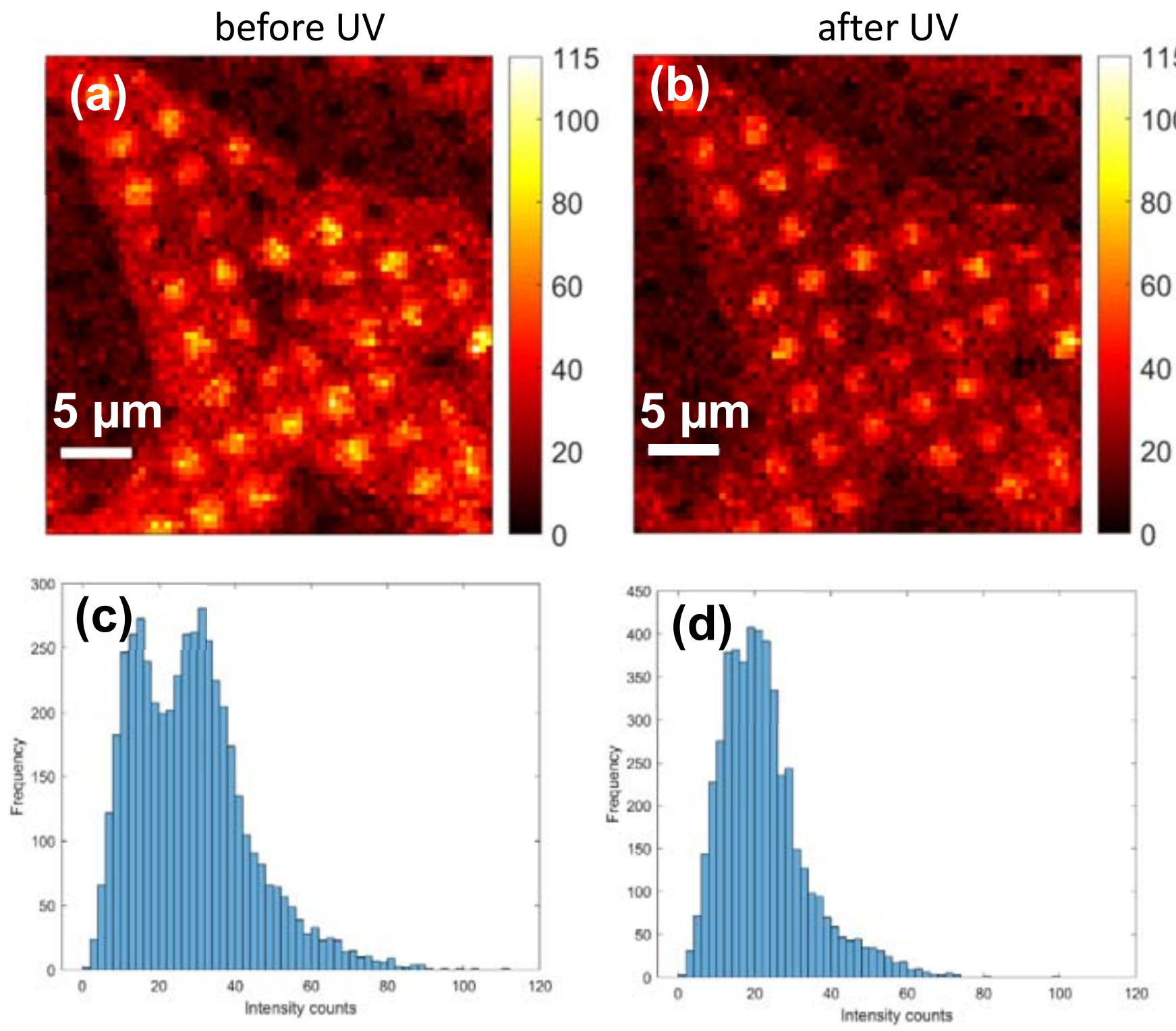

**Fig. S18:** PL maps acquired on a ML-$MoS_2$ crystal overlying on a Quantifoil type TEM grid before and after the UV exposure (365 nm, ~30 mW/cm$^2$ , 5 min) are shown in (a) and (b). Spatial maps of PL on pristine and UV exposed regions on the Quantifoil grid reveal a clearly visible quenching effect, however being quantitatively not of the same order of the quenching observed with UV exposed crystal on the $SiO_2$/Si substrate (shown in Fig. 5 of the main manuscript). In order to elucidate the quenching of PL, histograms of intensity levels from PL maps are extracted, as seen in (c), (d). Studying histograms of PL counts from the ML-$MoS_2$ before and after the UV exposure would reveal if there is any shift or reshaping of the distribution of counts to the lower values of counts after UV exposure. Clearly, it can be observed from the histograms that the fraction of frequencies of bins with counts exceeding 60 (corresponding dominantly to PL counts from the pristine ML-$MoS_2$ crystal situated on the holes of TEM grid) decreased significantly after the UV exposure. This observation infers that as a result of the UV exposure, the PL signal across the crystal overlaying on the TEM grid holes is quenched and thus causing a reshaping of the histogram.

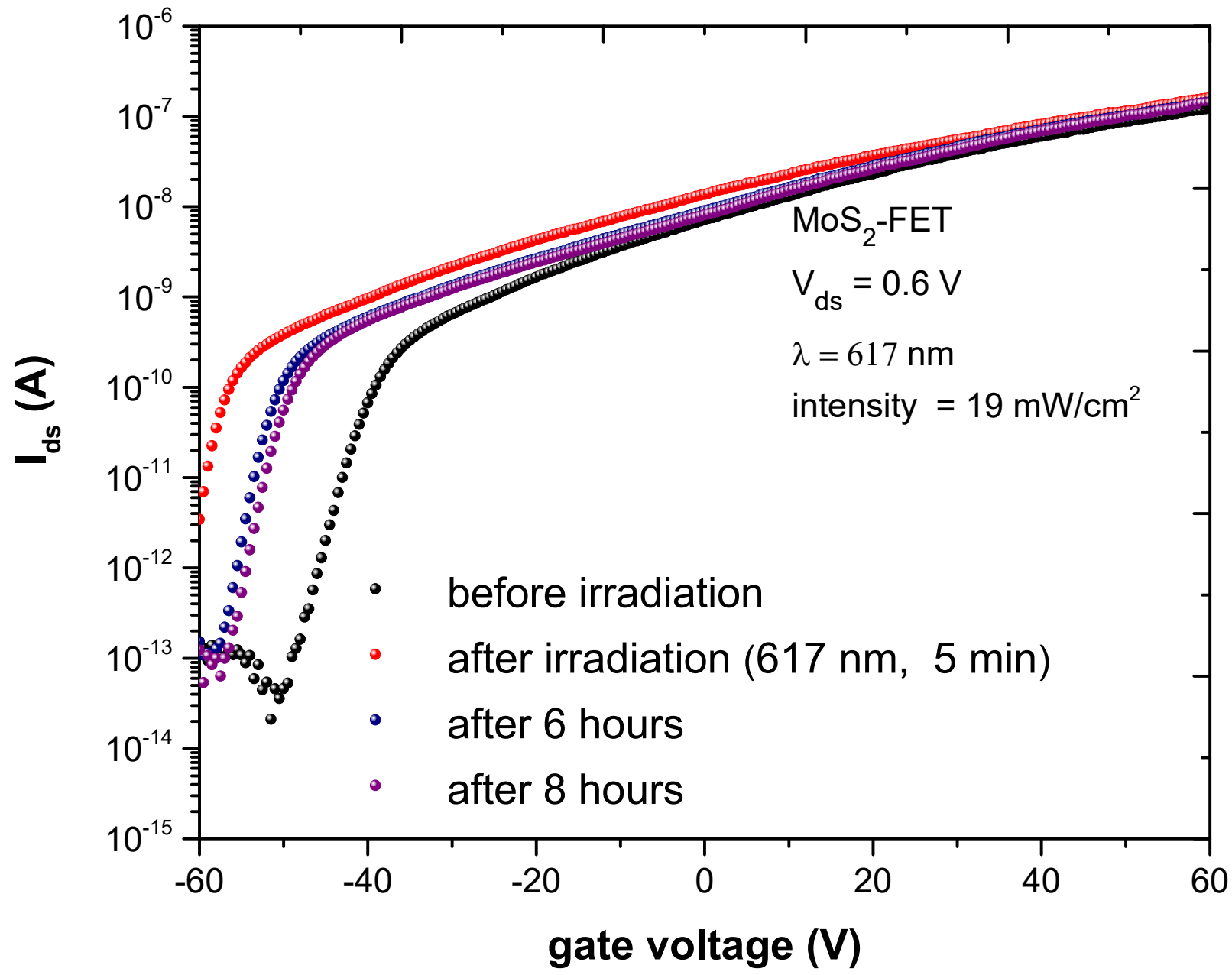

**Fig. S19:** Decay of the PPC effect in a $MoS_2$-FET after irradiation with light of wavelength 617 nm (intensity = 19 mW/cm$^2$) for 5 min. The device shows only a weak PPC comparing to the GPPC after UV irradiation.

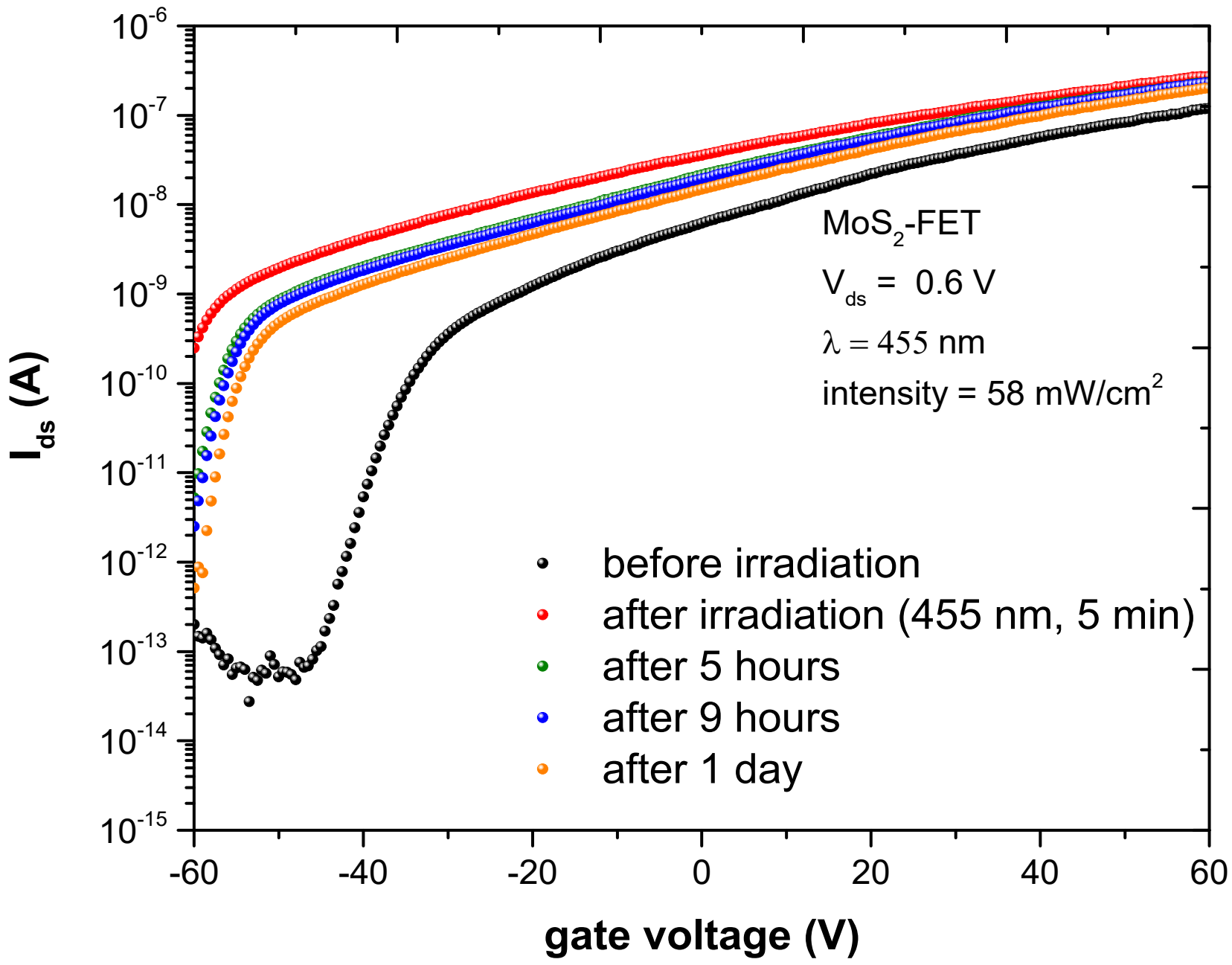

**Fig. S20:** Decay of the PPC in a $MoS_2$-FET after irradiation with light of wavelength 455 nm (intensity = 58 mW/cm$^2$) for 5 min. The device shows only a weak PPC effect comparing to the GPPC effect after UV irradiation.